\documentclass{scrartcl}

\KOMAoptions{paper=a4,paper=portrait,fontsize=10pt}
\usepackage[english]{babel}
\usepackage[ansinew]{inputenc}
\usepackage[T1]{fontenc}
\usepackage{helvet}
\usepackage[a4paper,portrait]{geometry}
\let\oldtwocolumn\twocolumn

\newif\iftwocolumn\twocolumntrue

\def\twocolumn{\twocolumntrue}
\usepackage{indentfirst}
\newlength{\OneColumnWidth}
\newlength{\TwoColumnWidth}
\makeatletter
\renewcommand\section{\scr@startsection{section}{1}{\z@}{-3.5ex \@plus -1ex \@minus -.2ex}{2.3ex \@plus.2ex}{\normalfont\bfseries}}
\renewcommand\subsection{\scr@startsection{subsection}{2}{\z@}{-3.5ex \@plus -1ex \@minus -.2ex}{2.3ex \@plus.2ex}{\normalfont\bfseries}}
\makeatother

\usepackage[
	colorlinks=true,
	urlcolor=blue, % \href{...}{...} external URL
	filecolor=blue, % \href{...} local file
	linkcolor=blue, % \ref{...} and \pageref{...}
	citecolor=blue, % \cite{...}
]{hyperref}
\usepackage{cite} % [1-3] instead of [1,2,3]

\usepackage[singlelinecheck=true,font=small,labelfont=bf,format=plain]{caption} % format: plain = no indent, hang = with indent

\let\oldhline\hline
\renewcommand{\hline}{\oldhline\rule{0pt}{12pt}}
\usepackage{enumitem}
\setlist{nosep}
\setenumerate[1]{label=(\arabic*)}
\setenumerate[2]{label=(\Alph*)}

\usepackage{ifthen}
\usepackage{forloop}
\usepackage{etoolbox}
\usepackage{graphicx}
\usepackage[fleqn]{amsmath}

\usepackage{tabularx} % Tabellen mit relativer Spaltenbreite
\newcolumntype{L}[1]{>{\raggedright\arraybackslash}m{#1}} % linksbündig mit vorgegebener Breite
\newcolumntype{R}[1]{>{\raggedleft\arraybackslash}m{#1}} % zentriert mit vorgegebener Breite
\newcolumntype{C}[1]{>{\centering\arraybackslash}m{#1}} % rechtsbündig mit vorgegebener Breite

\usepackage{xcolor} % Farben 
\usepackage{graphicx} % Zum Laden von Grafiken
\usepackage{wrapfig} % Textumflossene Bilder
\usepackage{tikz} % Bildmanipulation

\usepackage{amsmath} % AMS Math Package
\usepackage{amsthm} % Theorem Formatting
\usepackage{amssymb} % Math symbols such as
\usepackage{amsfonts} % Math fonts
\usepackage{amstext} % Math fonts
\usepackage{mathrsfs}
\usepackage{mathtools}
\usepackage{upgreek}
\usepackage{rotating}
\usepackage{verbatim} % f?r comment

\usepackage{multicol} % Allows for multiple columns
\usepackage{multirow} % Allows for multiple rows
\usepackage{threeparttable} % Tabellen mit Fu?noten
\renewcommand{\title}[1]{\def\inserttitle{#1}}
\newcommand{\email}[1]{\def\insertemail{#1}}
\renewcommand{\abstract}[1]{\def\insertabstract{#1}}
\def\insertjournal{}
\def\insertdoi{}
\def\insertarxiv{}
\newcommand{\journal}[3][accepted]{
	\def\tmpa{#1}\def\tmpb{submitted}\ifx\tmpa\tmpb\def\journalpre{Submitted to: }\else\def\tmpb{accepted}\ifx\tmpa\tmpb\def\journalpre{Accepted for publication in: }\else\def\tmpb{prepared}\ifx\tmpa\tmpb\def\journalpre{Prepared for submission to: }\else\def\journalpre{}\fi\fi\fi
	\if\relax\detokenize{#3}\relax\def\insertjournal{\journalpre#2}\else\def\insertjournal{\journalpre\href{#3}{#2}}\fi
}
\newcommand{\doi}[1]{\if\relax\detokenize{#1}\relax\def\insertdoi{}\else\def\insertdoi{DOI: \href{http://dx.doi.org/#1}{#1}}\fi}
\newcommand{\arxiv}[2]{\if\relax\detokenize{#2}\relax\def\insertarxiv{}\else\def\insertarxiv{arXiv: \href{https://arxiv.org/abs/#1}{#1 [#2]}}\fi}
\newcounter{authors}
\newcounter{addresses}
\newcounter{keywords}
\newcommand{\addauthor}[2]{\csdef{author\arabic{authors}}{#1}\csdef{authoraddress\arabic{authors}}{#2}\stepcounter{authors}}
\newcommand{\addaddress}[1]{\csdef{address\arabic{addresses}}{#1}\stepcounter{addresses}}
\newcommand{\addkeyword}[1]{\csdef{keyword\arabic{keywords}}{#1}\stepcounter{keywords}}
\newcommand{\maketitlesub}{
	\newcounter{i}
	\newcounter{j}
	\noindent\textbf{%
		\Large\inserttitle\\[1em]
		\large\csuse{author0}$^{\csuse{authoraddress0}}$%
		\forloop{i}{1}{\value{i} < \value{authors}}{%
			, \csuse{author\arabic{i}}$^{\csuse{authoraddress\arabic{i}}}$%
		}
	}\\[1em]
	\normalsize
	\setcounter{j}{0}
	\forloop{i}{0}{\value{i} < \value{addresses}}{%
		\stepcounter{j}
		\ifnum\value{addresses}>1$^{\arabic{j}}$\fi\,\csuse{address\arabic{i}}
		\ifthenelse{\value{j}<\value{addresses}}{\\}{}
	}
	\ifx\insertemail\empty\\[1em]\else\\[0.5em]E-mail address: \insertemail\\[1em]\fi
	\textbf{Abstract:} \insertabstract
	\ifthenelse{\value{keywords}=0}{}{
		\\[1em]
		Keywords: \csuse{keyword0}%
			\forloop{i}{1}{\value{i} < \value{keywords}}{%
				; \csuse{keyword\arabic{i}}%
			}
	}
}
\renewcommand{\maketitle}{\iftwocolumn\oldtwocolumn[\maketitlesub\vspace{1.5em}]\else\maketitlesub\fi}

\usepackage[headsepline]{scrlayer-scrpage}
\clearpairofpagestyles
\ihead{%
	\ifx\insertarxiv\empty%
			\ifx\insertjournal\empty\else\textnormal\insertjournal\fi%
	\else%
		\ifx\insertdoi\empty%
			\ifx\insertjournal\empty\else\textnormal\insertjournal\fi%
		\else%
			\ifx\insertjournal\empty\else\textnormal\insertjournal\fi%
			\ifx\textnormal\empty\else\linebreak\textnormal\insertdoi\fi%
		\fi%
	\fi%
}
\ohead{%
	\ifx\insertarxiv\empty%
		\ifx\textnormal\empty\else\textnormal\insertdoi\fi%
	\else%
		\ifx\insertdoi\empty%
			\ifx\insertarxiv\empty\else\textnormal\insertarxiv\fi%
		\else%
			\ifx\insertarxiv\empty\else\linebreak\textnormal\insertarxiv\fi%
		\fi%
	\fi%
}
\ifoot{}
\ofoot{}

\usepackage[hang]{footmisc}
\let\oldthebibliography\thebibliography

\renewcommand\thebibliography[1]{
	\small
	\oldthebibliography{#1}
	\setlength{\parskip}{0pt}
	\setlength{\itemsep}{0pt plus 0.3ex}
}
\usepackage{blindtext}

\twocolumn % \onecolumn || \twocolumn

\title{Coupled deformed microdisk cavities featuring non-Hermitian properties}

\addauthor{Tom Rodemund}{1}
\addauthor{S\'ile Nic Chormaic}{2}
\addauthor{Martina Hentschel}{1}

\addaddress{Institute of Physics, Technische Universit\"at Chemnitz, 09107 Chemnitz, Germany}
\addaddress{Okinawa Institute of Science and Technology Graduate University, Onna, Okinawa, Japan}

\email{martina.hentschel@physik.tu-chemnitz.de}

\abstract{
Coupled cavities are of interest as they expose qualitatively new effects, such as non-Hermitian properties, that are \resubm{beyond the possibilitie} of %not accessible using an 
individual cavities.
\resubm{Here, we investigate the coupling between two dielectric two-dimensional microdisk cavities and compare circular vs.~deformed (lima\c{c}on) resonator shapes 
	\comments{are circular and deformed separate or are you saying the deformed is limacon plus circular at the same time?  If the latter, I'd say "deformed circular" shape} 
	\commentm{Solution: compare circular vs.~deformed (lima\c{c}on) resonator shapes }
	as a function of their distance and address the effect of coupling on the far-field emission properties. We find that the asymmetric coupling characteristic for non-circular, deformed cavities  induces 
	non-Hermitian properties prominently evident in a mode-dependent chirality of the coupled cavity modes.}
We use an analytical model to explain our findings and reveal the direct connection between coupling asymmetry and the resulting sense of rotation of the coupled modes. 
\resubm{While the overall far-field directionality remains robust for intercavity distances larger than two wavelengths, we observe enhanced and reversed emission for smaller distances even for only two coupled cavities.}
Our findings could prove useful for future applications such as far-field emission control \resubm{and sensing}. %of coupled cavities.
}

\addkeyword{limacon}
\addkeyword{optical microcavity}
\addkeyword{non-Hermitian system}
\addkeyword{phase-space}
\addkeyword{Husimi function}
\addkeyword{directional emission}
\addkeyword{whispering-gallery mode}

\journal[submitted]{Applied Physics Letters}{}
\newcommand{\commentm}[1]{} % comment command for martina
\newcommand{\comments}[1]{} % comment command for sile
\newcommand{\commentt}[1]{} % comment command for tom
\newcommand{\resubm}[1]{\textcolor{black}{#1} } % comment command for tom

\newcommand{\imag}{\mathrm{i}}

\newcommand{\lima}{lima\c{c}on }

\newcommand{\ocoupl}{\delta} % o-coupling
\newcommand{\ocouplu}{\ocoupl_\mathrm{u}} % up
\newcommand{\ocoupld}{\ocoupl_\mathrm{d}} % down

\newcommand{\xcoupl}{\kappa} % x-coupling
\newcommand{\xcouplu}{\xcoupl_\mathrm{u}} % up
\newcommand{\xcoupld}{\xcoupl_\mathrm{d}} % down

\newcommand{\sectiontitle}[1]{}

\newcommand{\onlinecite}[1]{\cite{#1}}

\begin{document}

\maketitle

\sectiontitle{Introduction.}
Optical microcavities \cite{Vahala_book} have received a lot of attention over the past decades with an enormous application range from microlasers \cite{taka_susumu_2dlaser,7414429, Yu:20} and frequency combs \cite{PhysRevLett.101.093902, kippenberg2011microresonator, Yang:16, Chen2020, Tian:22},
% \comments{the references used were not for frequency combs so I've updated them}
through mesoscopic model systems arcing quantum chaos 
%\comments{I've added brackets around the "quantum" since maybe these papers are not all quantum?}
%\commentm{quantum is correct} 
\cite{QuaChaCav:Nockel:1997, gmachl_science, annbill, RMP_cao_wiersig, Sciencechaosforcoupling}, to sensing applications\cite{krioukov2002sensor, chen2017exceptional, yu2021whispering, franchi2023infinity}. 
%\comments{if you need to reduce length then remove the sensing application and the 3 associated references}
Key theoretical concepts originate from semiclassical approaches, including the ray-wave correspondence and, in particular, phase-space methods. 
In this paper, we will extend the system class to two coupled, passive optical microcavities \cite{ChilMinKim_Coupled,RyuHentschel_coupled2011,coupled_PRR2018} and use phase-space methods, such as Husimi functions, to  \resubm{illustrate the presence of non-Hermitian physics induced by the coupling of two asymmetric cavities, and assess} 
\comments{I guess it should be assess not access}
\commentm{of course}%increase 
their potential for future applications \resubm{such as far-field engineering}.

Notably, non-Hermitian physics \cite{Bender_2007, harari2018topological, ozawa2019topological} has increased our insight in a number of fields, including photonic crystals \cite{natcomm_coupphotcryst}  and optical microcavities \cite{nat_lanyang}.  Here, we introduce coupled optical microcavities as another model system \resubm{that was previously studied in the context of intercavity-distance dependent far-field properties \cite{coupled_PRR2018}. We focus on two coupled resonators and find %can be ruled by
	non-Hermitian effects to strongly alter the electromagnetic fields especially outside the resonators in comparison to the single cavity case. %influence 
	\comments{***snc: DO YOU MEAN THAT THE EFFECTS INFLUENCE THE SYSTEM PROPERTIES?****} 
	%the system properties.
	\commentm{Solution: replace "influence system properties" by  to strongly alter the electromagnetic fields especially outside the resonators in comparison to the single cavity case.}
}
%are important and can be illustrated using } %their mode of action using}
%Although this may initially appear unexpected,
%%While this might come as a surprise at first sight, 
%we show that 

The Letter is organized as follows. 
%In Section \ref{sec_model}, 
We first introduce our model system, which consists of two coupled, deformed \resubm{dielectric} microdisk cavities, and discuss the mode classification. We then 
%of the coupled system that is ruled by the presence of one axial symmetry axis (located midway between the two identical cavities). 
%In Section \ref{sec_regimes}, we 
vary the distance between the cavities and characterize the resulting coupling regimes. We show that the geometric asymmetries of the system imply asymmetric coupling and thus chirality as a generic feature of non-Hermitian physics. We analyze this behavior in phase-space using Husimi functions \cite{husimiEPL} and support our findings by analytical modeling.
%and we investigate this further in Section \ref{sec_nonHerm}. We introduce the Husimi function as a suitable tool to quantify the chirality of the system and support our findings by an analytical model in Section \ref{sec_anal}.
%, and discuss its dependence on the intercavity distance. 
%We extend our consideration to the far-field emission properties and find that the emission characteristics of coupled microdisks is highly dependent on their separation distance, and that an inversion of the typical emission directionality, previously observed in Ref.~\cite{coupled_PRR2018} for a chain of coupled cavities, can already be achieved for two coupled cavities. 
%We end with conclusions and an outlook.

%\commentt{This is a comment from Tom. The textwidth is \the\textwidth, the font is \expandafter\string\the\font}
%\commentm{This is a comment from Martina.}
%\comments{This is a comment from Síle.}

\begin{figure}[h]
\centering
\includegraphics[width=.45\textwidth]{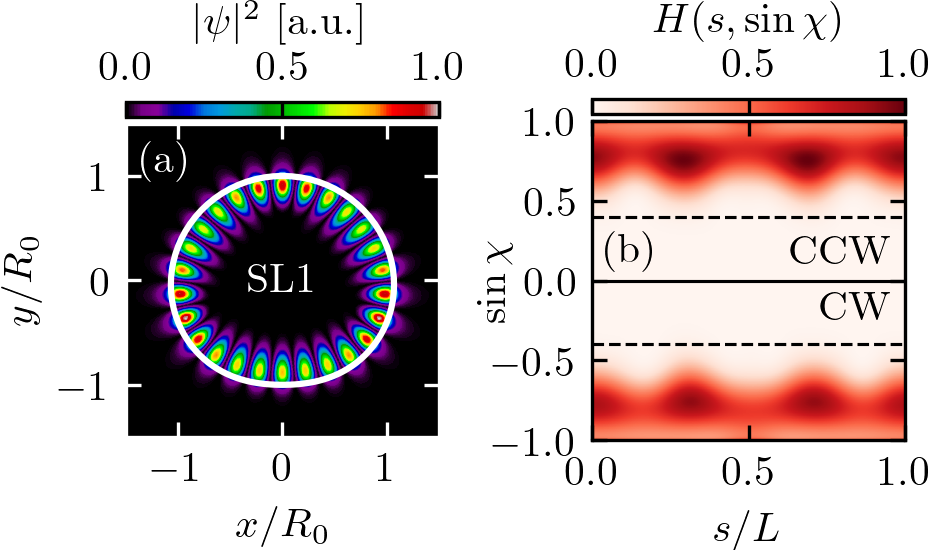}
\caption{
(a) Mode pattern and (b) Husimi function of the SL1 mode (see text for details).
The white line represents the cavity boundary.
The dashed line in (b) % in the Husimi function shows 
indicates the sine of the critical angle $|\sin \chi_\mathrm{crit}| = 1 / n$ with positive (negative) $\sin \chi$ indicating counterclockwise, CCW (clockwise, CW) motion.
% , where $n$ is the refractive index of the cavity.
%For an angle of incident $\chi > \chi_\mathrm{crit}$ total internal reflection is given.
The Husimi function shows the WG mode characteristics as it is confined to the region with total internal reflection outside the critical lines.
%    The three maxima correspond to the three positions at the boundary with $s/L \in \{0,0.3,0.7\}$, where the field intensity is highest.
% \comments{I think you should have explained L1 mode before the figure or do so in caption.  For the figure, the sin chi at the left hand axis is too close to the mode pattern, so it looks as though it is part of that figure rather than the Husimi.  I'd also add something in caption about Husimi... along the lines of "in the Husimi function we see the characteristic... that represents the modes in a WGM cavity}
% \commentt{Note that the Husimi function is slightly asymmetrical for negative and positive $\sin \chi$ due to numerical reasons, the Husimi function is very sensitive to this.}
%   \commentt{Implemented comments.}
}
\label{fig:mode_L1}
\end{figure}

\begin{figure}[h]
\centering
\includegraphics[width=.45\textwidth]{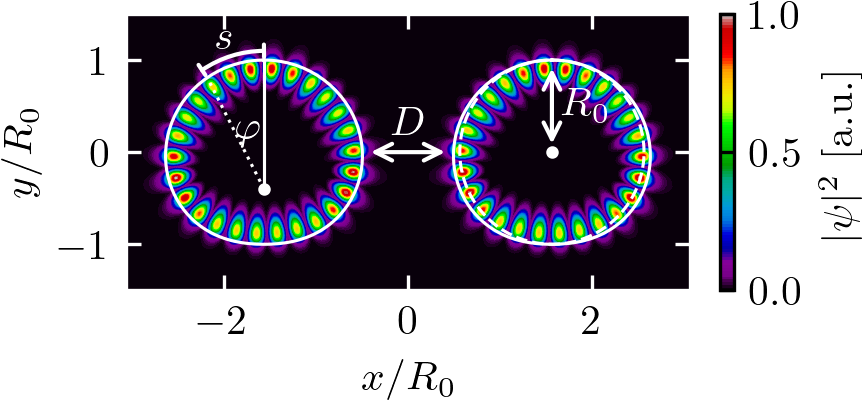}
\caption{Model system consisting of two \lima microcavities. The even CL1 mode (see text for details) intensity distribution  is shown at an intercavity distance $D/\lambda=1$ (\resubm{transient} coupling regime, the modes are very similar to the single cavity modes, here SL1).
%The parameters used are introduced. 
%, which are defined by eq.~\ref{eq:limacon} (see left cavity for definitions). A \lima resembles a slightly deformed disc with radius, $R_0$ (see right cavity where the boundary of the corresponding disc is shown by the dashed white line). 
%The cavity boundaries are separated by a distance $D$.
}
\label{fig:modelsystem}
\end{figure}

\begin{figure*}
\centering
\includegraphics[width=0.8\textwidth]{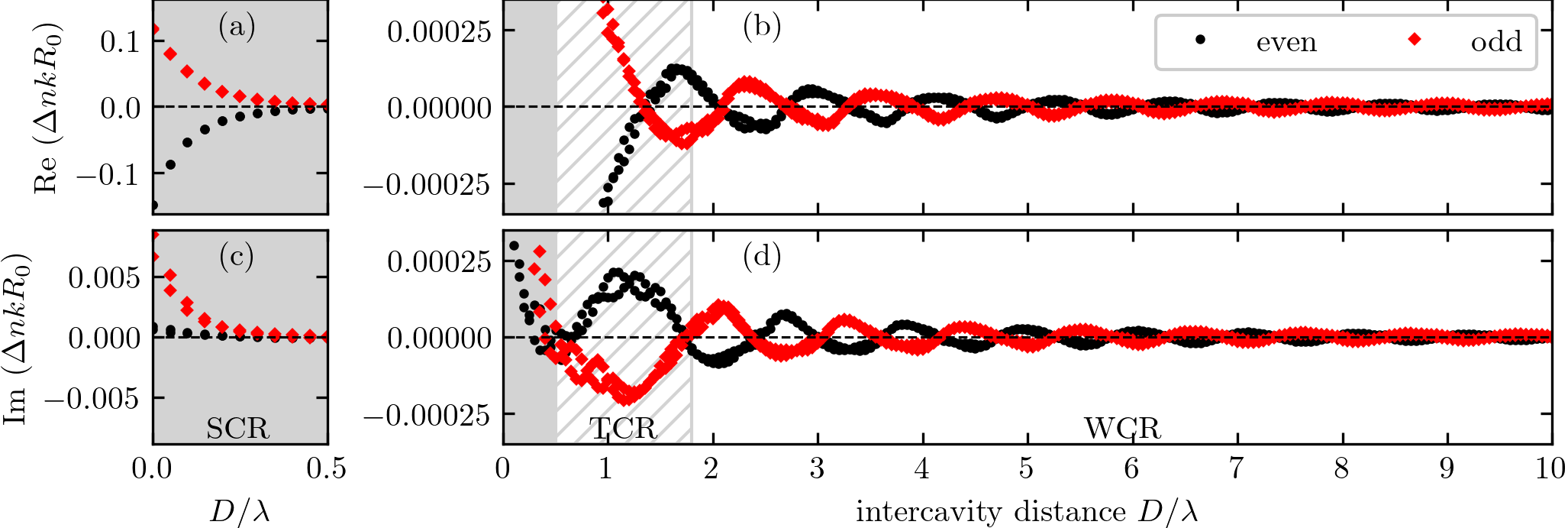}
\caption{%\resubm{CHECK Add TCR} \commentt{Done!}
Complex eigenvalues of the coupled system's \resubm{CL1 mode with even (black dots) and odd (red diamonds) intercavity parity (along the $x=0$ axis)} as function of the intercavity distance $D/\lambda$. \resubm{Each of these two curves consists of two symbols for each $D/\lambda$ representing the even and odd intracavity parity.}
%namely symmetric (or even, black dots) and antisymmetric (or odd, red diamonds) modes with respect to the intercavity symmetry axis and arbitrary intracavity symmetry. 
The deviation of the coupled system eigenvalue from the SL1 values for the real (imaginary) part in the upper (lower) panels is shown. The left (right) panels show the strong (weak) coupling regime named SCR (WCR). \resubm{The onset of the transient coupling regime (TCR) is characterized by larger deviations and indicated in the hatched gray region.}
%  Two coupling regimes can be distinguished: The strong-coupling regime (SCR) for $D/\lambda < 0.5$ (see left figures and gray area in right figures) and the weak-coupling regime (WCR) for $D/\lambda > 0.5$.
%  In the SCR the symmetric and antisymmetric values split significantly. 
%  In the WCR the eigenvalues oscillate with decreasing amplitude, exhibiting $\Delta n k R_0 \propto \exp(\imag k D) / D$ behavior.
}
\label{fig:array_eigenvalues}
\end{figure*}

%\section{Model system and mode characterization}
%\label{sec_model}

\sectiontitle{Model system and mode characterization.}
Our optical microcavity of  choice is the \lima cavity that has previously been studied in the context of directional emission from optical microcavities and microlasers \cite{limacon,limacon_Cao2009,limacon_Kim2009,limacon_Susumu_Taka2009,limacon_NJPhys,limacon_Yan2009,limacon_Albert2012,3dlimacon}. As in these works, we model it as a two-dimensional system.
%(which is justified for a system height on the order of a wavelength \cite{coupled_PRR2018}). 
In polar coordinates $(R, \varphi)$ the cavity's shape is given as
%\begin{equation} \label{eq:limacon}
$  R(\varphi) = R_0 (1 + \varepsilon \cos \varphi) $
%\end{equation}
where  $R_0$ is the mean radius and $\varepsilon$ is the deformation parameter. We use $\varepsilon=0.4$, which is a typical value with \resubm{increased} %pronounced %due to the high resulting 
far-field emission directionality\cite{limacon_Susumu_Taka2009}. The refractive index $n$ of the cavity is set to $n=2.5$, and $\lambda = 2 \pi/k$ is the wavelength in vacuum surrounding the cavity. %(we assume refractive index of 1 outside the cavity). 

A characteristic, whispering-gallery (WG) type mode pattern for a high quality ($Q$) factor  \lima cavity is shown in Fig.~\ref{fig:mode_L1}(a).
We see that the WG-type field has higher intensity $|\psi|^2$ in some regions than others due to the non-circular geometry of the cavity.
This is reflected in its Husimi function, a phase-space representation of the wave pattern \cite{husimiEPL},  in Fig.~\ref{fig:mode_L1}(b) where the darker regions correspond to regions with higher field intensity (see below for more details).
The phase space is spanned by the arc length $s$ along the circumference (with total length $L$) of the cavity, cf.~Fig.~\ref{fig:modelsystem}, and the sine of the angle of incidence $\chi$.
%\comments{here I'd add some comment on the pattern observed, such as the fact that the fields are stronger in some regions of the cavity than others, due to the non-circular geometry of the cavity }
%The associated Husimi function, \textcolor{red}{which represents the...} is given in Fig.~\ref{fig:mode_L1}(b).  We will discuss the Husimi function further in Section \ref{sec_nonHerm} \commentm{check this makes sense}.  
We use the COMSOL software package to compute the electromagnetic modes of both the single cavity and coupled cavity system. Here, we focus on modes with out-of-plane electric field components, i.e.~the typically studied TM modes.

Specifically we examine the following modes:
Two modes of the single \lima cavity,
%, where the unperturbed system has 
with $nkR = 15.91$ and \resubm{quality factor} $Q = 1.60 \times 10^4$, and $nkR = 18.04$ with $Q = 5.55 \times 10^4$, which we refer to as SL1 and SL2. 
The \resubm{correponding} coupled system modes %with frequencies close to SL1 and SL2 
are named CL1 and CL2. 
%This distinction needs to be made, as the coupling results in a shift of the eigenvalues and other qualitatively different behavior to the single system.
A\resubm{circular} microdisc without deformation is also investigated, and the mode with $nkR = 16.56$ and $Q = 8.53 \times 10^5$ is referred to as SD and CD for the single cavity and coupled system, respectively. 

%\comments{maybe state why you focus on these modes}
%cf.:  We will consider TM % TE CHECKREV
%polarized light (electric field transverse to the resonator plane, i.e. along %the $z$ axis, $\Vec{E} = E_z \Vec{e_z}$),

%\begin{itemize}
%    \item use COMSOL
%    \item use out-of-plane electric fields
%\end{itemize}

The coupled system is shown in Fig.~\ref{fig:modelsystem} with the symmetric coupled-system mode CL1.
The system parameters, i.e., intercavity distance $D$, mean single cavity radius $R_0$, arc length $s$, and polar angle $\varphi$, are indicated.
Most notably, the extension to two cavities brings in a new symmetry axis, \resubm{namely the inter-cavity} mirror symmetry axis %in between the two cavities that lies
lying parallel to the \resubm{intra-cavity} mirror reflection axes of each single cavity.
This induces a new quantization, or classification, condition: the electric field \resubm{amplitude} on the coupled system axis has to be extremal or has to vanish.
We classify the corresponding modes as symmetric (even) or antisymmetric (odd), respectively.
Two modes exist for each of these two classifications (i.e., four modes in total)
\resubm{since the symmetry classification with respect to the intracavity mirror symmetry axis can be either symmetric (even) or antisymmetric (odd) as well.} 
%and they differ in their array symmetry properties, being (anti)symmetric with respect to the intercavity symmetry axis.

%\section{Coupling regimes}
%\label{sec_regimes}

\sectiontitle{Coupling regimes.}
The evolution of the coupled system eigenmodes as a function of the intercavity distance $D$ is shown in Fig.~\ref{fig:array_eigenvalues} where the deviations of the real and imaginary parts of the eigenmodes from the respective single cavity modes are shown.  
The most prominent feature is an oscillatory behavior  for \resubm{ $D/\lambda > 1.5$} %$D/\lambda > 0.5$ 
with regular oscillations damping out for larger $D$,
%In the WCR the eigenvalues oscillate with decreasing amplitude, 
exhibiting a $\Delta n k R_0 \propto \exp(\imag k D) / D$ behavior.
We refer to this as the weak coupling regime (weak CR or WCR). The oscillations originate from the quantization requirement along the intercavity axis. This can be fulfilled by slightly expelling the mode out of (or into) the cavity, causing a change in the imaginary part in $nkR$ (and, consequently, in the $Q$ factor), or by adjusting the wavelength, i.e., changing the real part of $nkR$. Both mechanisms alternate and yield the out-of-phase oscillations in the real and imaginary parts, see Fig. 1 in the Supplementary Material (SM) for details.

A qualitatively different behavior is seen for $D/\lambda < 0.5$, cf.~the left panels in Fig.~\ref{fig:array_eigenvalues}. In this strong coupling regime (strong CR or SCR) 
%(see left figures and gray area in right figures) and the weak-coupling regime (WCR) for $D/\lambda > 0.5$
the symmetric and antisymmetric modes do not oscillate any more but rather split significantly, indicating a different interaction of the coupled cavities. 
Similar values for the onset of the strong coupling regime were found for two coupled disk cavities\cite{ChilMinKim_Coupled} %studied in detail in Ref.~\cite{ChilMinKim_Coupled} 
and for a linear array of cavities \cite{coupled_PRR2018}.   \resubm{Between the strong and weak coupling regimes we see a transient behavior (transient CR or TCR) with larger oscillations and application potential as we discuss in the following.}
%CHECK: HIER VERWEIS AUF PAPER CHIL MIN KIM!!! 

%\begin{itemize}
%    \item discuss Fig.~\ref{fig:array_eigenvalues} (central point)
%    \item regimes are closely intertwined with everything, thus hard to seperate from the rest
%    \item reference to chapter 4, where it is discussed in more detail
%\end{itemize}

%\section{Chirality and non-Hermitian properties: A phase-space study}
%\label{sec_nonHerm}

\begin{figure*} % show this first, use as guidance for all
\centering
\includegraphics[width=15cm]{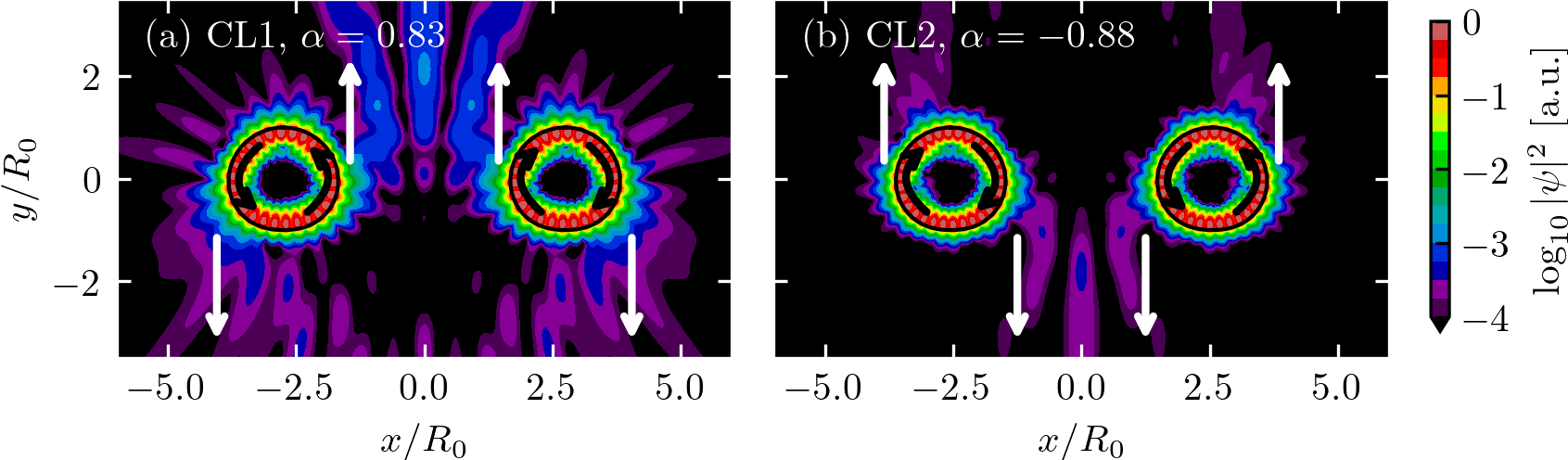}
\caption{ 
Logarithmic field intensity $\log |\psi|^2$ of the two even (symmetric) coupled-system modes (a) CL1 and (b) CL2 at distance $D/\lambda = 3.4$. 
The black arrows inside the cavities indicate the prevalent traveling wave being (a) CCW and (b) CW in the left cavity and yielding positive and negative chirality $\alpha$, respectively. The right cavity shows the opposite rotation. 
The arrows outside indicate the associated major emission directions %show qualitatively the location and direction of emission 
of the cavities. The dependence on chirality is evident. 
%Both modes exhibit a strong rotation.
% \comments{should you add a scale colour bar?}
% \commentt{Yes, it is there now. I chose a low number of levels to reduce the clutter in the image.}
%For CL1, a predominantly CCW rotation is found in the left cavity.
%For CL2, a dominance of CW traveling-waves is apparent.
%\comments{are cw and CCW defined?}
%\commentt{I added it at the beginning of Sec.~\ref{sec_nonHerm} as it is needed there first.}
%Depending on the mode rotation, the emission between the cavities has its intensity maximum at $\theta = 0$ or $\theta = \pi$.
%The behavior of the right cavity is given by symmetry.
}
\label{fig:array_modes}
\end{figure*}

\begin{figure*}
\centering
\includegraphics[width=15cm]{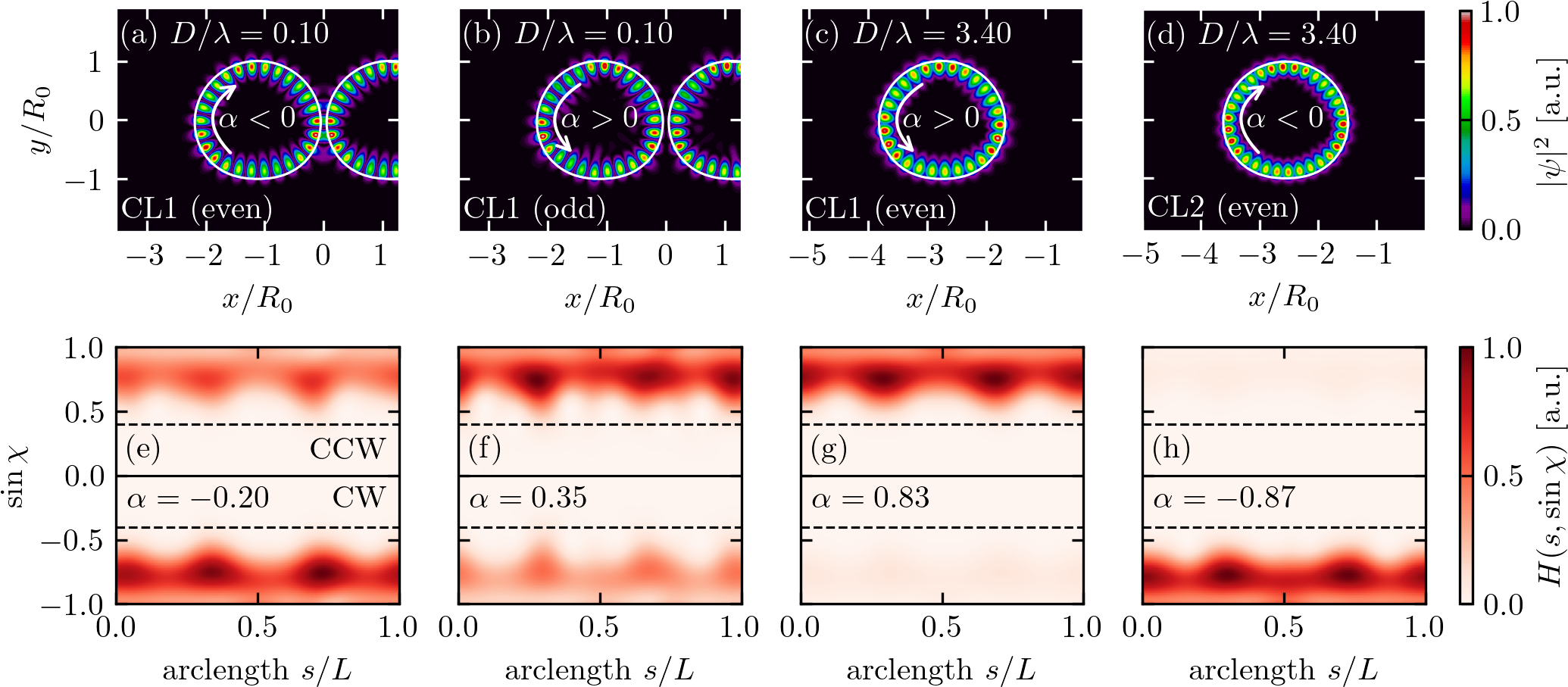}
\caption{
Electric field intensity $|\psi|^2$ (a,b,c,d; color scale as in Fig.~\ref{fig:modelsystem}) and corresponding Husimi function of the left cavity (d,e,f,g, color scale as in Fig.~\ref{fig:mode_L1}) for various intercavity distances, parities, and frequencies as indicated in the figures; same color scale  as in 
%For the values corresponding to the colors see 
Figs.~\ref{fig:mode_L1} and \ref{fig:modelsystem}.
%The arrows in the field plots mark the 
The dominant WG-type wave propagation direction is marked by the white arrow in (a,b,c,d).
The dashed lines in the Husimi plots (e,f,g,h) mark the critical angle $\sin \chi_\mathrm{crit} = 1/n$. %, below which total internal reflection is no longer occurring.
The solid line at $\sin \chi = 0$ separates the CW and CCW contributions of the Husimi function (see Eqs.~(\ref{eq:chirality}) and (\ref{eq:chirality_parts})). Different weight in the CW and CCW Husimi components signifies a finite chirality (measured with respect to the left cavity) that is induced by asymmetric coupling to right cavity and can visually be identified by blurred nodes in the intensity pattern as in (d). 
%    The emerging chirality is evident, where the CW and CCW contributions are no longer identical due to the presence of the second cavity (compare to Fig.~\ref{fig:mode_L1}).
%The chirality $\alpha$ is a function of all four parameters mentioned above. For color scales of the fields see Fig.~\ref{fig:modelsystem}, for the Husimi function see Fig.~\ref{fig:mode_L1}.
}
\label{fig:husimi_lima}
\end{figure*}

\begin{figure}
\centering
\includegraphics[width=0.45\textwidth]{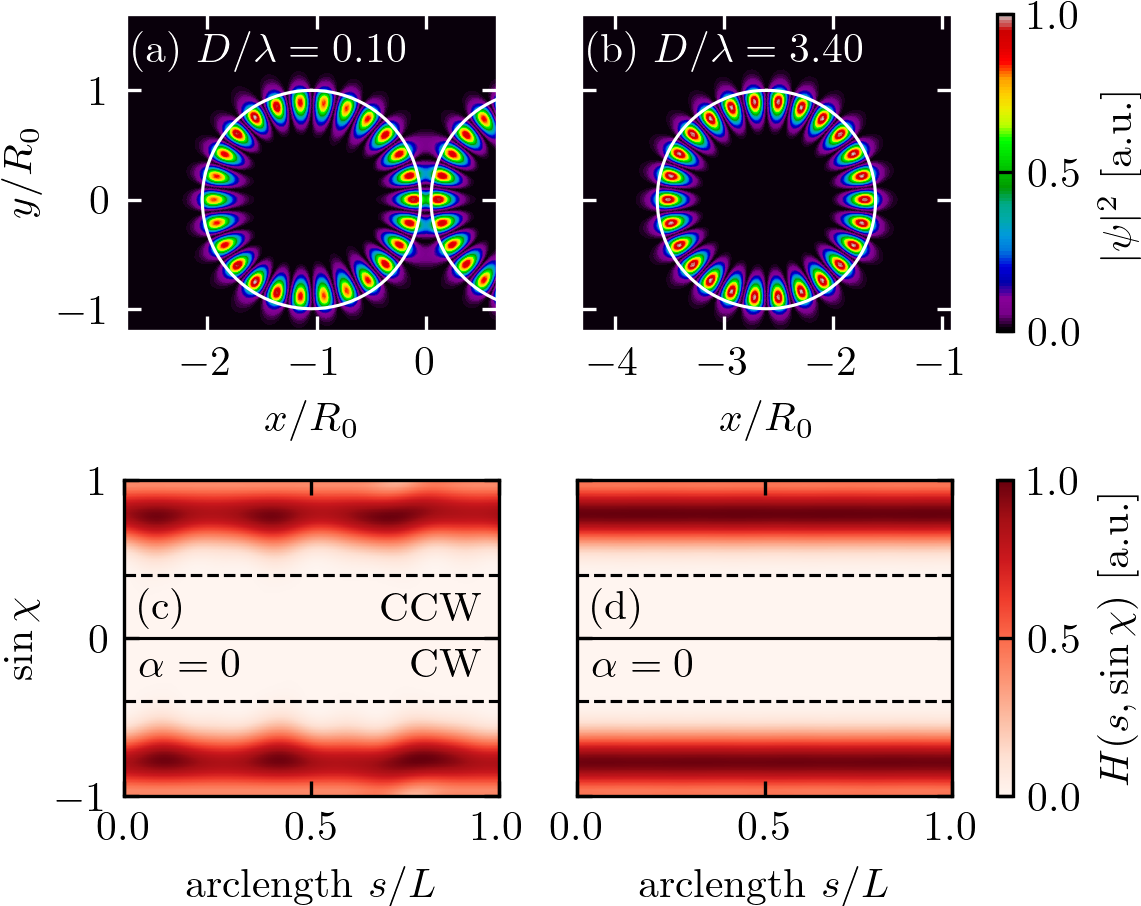}
\caption{
(a,b) Electric field intensity $|\psi|^2$  and (c,d) corresponding Husimi function (of the left cavity) of two coupled ideal disk-cavities at  different distances $D$.
Two even configurations of the CD mode are shown.
%(no deformamicrodiscs without deformation. %\commentm{CITE KIM ET AL HERE AND DISCUSS IN TEXT}
In the weak CR %WCR 
%  \comments{is WCR defined?} \commentt{Yes, in Sec.~\ref{sec_regimes}}
(b,d) the mode resembles an unperturbed WG mode, %identical to the unperturbed case.
whereas in the strong CR 
% \comments{is scr defined?}
(a,c) %(see left column) 
its Husimi function reveals some structure in phase space. 
%and wave function is visibly deformed.
However, due to the high symmetry of the system 
%along both the $x$ and $y$ axes, 
no chirality is present.
}
\label{fig:husimi_disc}
\end{figure}

Chirality is a key feature of non-Hermitian systems \cite{RMP_cao_wiersig}. It can be quantified by means of the non-orthogonality of counterpropagating CW (clockwise) and CCW (counter-clockwise) WG-type mode pairs and has been studied in (deformed) microdisks \cite{nonorth_asymm,sarma_ge_wiersig,shim_kim_expoint2009} and spiral cavities \cite{nonorth_spiral}. As the counterpropagating WG (type) modes gain different weight, a chirality, i.e.~a preferred sense of rotation, emerges.

Characteristic examples of such chiral modes are shown in Fig.~\ref{fig:array_modes} for the two different symmetric coupled-cavity modes CL1 and CL2. 
It illustrates that both chiralities can be readily realized in a coupled system. In  Fig.~\ref{fig:array_modes}(a), the coupled system mode CL1 
%originates from the single cavity mode SL1 and 
yields a positive chirality $\alpha$ in the left cavity, in contrast to the CL2 mode
%SL2-mode based resonance 
with opposite chirality. % with opposite sign. 
%CHECK RESUBM Taken out: 
%Taking into account the non-isotropic, directional emission characteristics of the \lima cavity, it is evident that the chirality has a strong influence on the far-field emission properties. Such an effect, including emission reversal, was previously observed in asymmetric cavities \cite{sarma_ge_wiersig} and linear cavity arrays \cite{coupled_PRR2018}.
%For the system of two coupled cavities studied here the arising chirality 
This is evident in the Husimi functions \cite{husimiEPL} of the system that possess %and observing 
a distinctly different weight in the CCW and CW contributions, cf.~Fig.~\ref{fig:husimi_lima}. This does not occur for two coupled disk cavities as shown in Fig.~\ref{fig:husimi_disc}.

%\subsection{Husimi functions for coupled cavities}

%\begin{itemize}
%    \item show Husimis and wave functions for lima:
%        \begin{itemize}
%            \item 2x $D/\lambda = 0.1$ (even, odd)
%            \item 2x $D/\lambda = 3.4$ (low $k$, high $k$)
%        \end{itemize}
%    \item show Husimis for disc (less so it gets less weight):
%        \begin{itemize}
%            \item 1x $D/\lambda = 0.1$ (even or odd)
%            \item 1x $D/\lambda = 3.4$
%           \item single file in text
%       \end{itemize}
%\end{itemize}

\sectiontitle{Chirality and non-Hermitian properties: A phase-space study.}
Here, we introduce chirality $\alpha$ via the Husimi function
%and study it as a function of distance, cf.~Fig.~\ref{fig:dir_and_chir} below. 
%To this end we introduce Husimi functions 
for single \cite{husimiEPL} and coupled \cite{Husimi_Bosch:21} cavities.
The Husimi function $H(s,\sin \chi)$ is based on the real space wave function  $\psi(s)$ and its normal derivative $\psi'(s)$ at the (inner) cavity boundary $s$ \cite{husimiEPL,SchomerusHentschel_phase-space, Husimi_Bosch:21}. 
%and can be deduced from it \cite{husimiEPL,SchomerusHentschel_phase-space, Husimi_Bosch:21}. %While four Husimi functions corresponding to incoming and outgoing beams inside and outside the cavity can
%a phase-space function, which corresponds to the surface-of-section of specular billards in ray dynamics.
We focus on the Husimi function $H(s,\sin \chi)$ corresponding to incoming  waves inside the cavity defined as \cite{husimiEPL} 
%it is defined as \comments{add citation}
\begin{equation} \label{eq:husimi}
H(s,\sin \chi) = \frac{n k}{2 \pi} \left|-\mathcal{F}(\chi) h(s,\sin \chi) + \frac{\imag}{k \mathcal{F}(\chi)} h'(s,\sin \chi)\right|^2
\end{equation}
with $\mathcal{F}(\chi) = \sqrt{n \cos \chi}$ and the overlap functions $h, h'$ given by
\begin{subequations}
\begin{align}
h(s,\sin \chi) &= \oint d s' \psi(s') \xi (s'; s,\sin \chi) &\quad \text{and}\\
h'(s,\sin \chi) &= \oint d s' \psi'(s') \xi(s'; s,\sin \chi) \:.&
\end{align}
\end{subequations}
%
%where $\psi, \psi'$ are the wave function and its normal derivative at the (inner) boundary point $s'$.
$\xi(s'; s,\sin \chi)$ is a minimal uncertainty wave packet with $\sigma = \sqrt{2}/(nk)$ centered around $(s', \sin \chi)$, which is defined as
\begin{eqnarray}
&&\xi(s'; s,\sin \chi) = \nonumber (\sigma \pi)^{-\frac{1}{4}} \\ 
&&\times \sum_{j\in \mathbb{Z}} \exp \left[ \frac{-(s'-s+jL)^2}{2\sigma} - \imag n k \sin(\chi) (s'-s + jL) \right]\:.
\end{eqnarray}
%\begin{align}
%    &&\xi(s'; s,\sin \chi) = \nonumber \\ 
%    &&(\sigma \pi)^{-\frac{1}{4}} \Sigma_{l\in Z} \exp \left[ \frac{-(s'-s+2\pi l)^2}{2\sigma} - i n k \sin(\chi+2 \pi l) \right]\:.&&
%\end{align}

Characteristic (incoming, incident) Husimi functions for various modes and intercavity distances of the two-\lima system are shown in Fig.~\ref{fig:husimi_lima}. 
The unbalanced contribution of CCW and CW components is evident. 
%RESEBM TAKEN OUT WAR SCHON 
%We point out that this is not the case for coupled disk cavities where CCW and CW mode contributions are balanced as shown in Fig.~\ref{fig:husimi_disc}. 

%\subsection{Chirality and farfield-emission directionality}

%\blindtext
We define the chirality $\alpha$ using the Husimi functions as 
%it distinguishes the contributions of CW and CCW modes
%\comments{define your acronyms}. 
%We define it as
%
\begin{equation} \label{eq:chirality}
\alpha = \frac{\alpha^+ - \alpha^-}{\alpha^+ + \alpha^-}
\end{equation}
with
\begin{equation} \label{eq:chirality_parts}
\alpha^\pm = \left| \int_0^{\pm 1} \int_0^L H(s,\sin \chi) \, \mathrm{d} s \, \mathrm{d} \sin \chi \right| \quad ,
\end{equation}
which yields $\alpha \in \left[-1,1\right]$ with $\alpha > 0 (<0)$ indicating dominance of CCW (CW) contributions in the left cavity.
% while $\alpha < 0$ is shows dominating CW contributions.

%CHECK: IN Footnote maybe?:
%Note that chirality is also often defined by expansion of the wave function in cylindrical harmonics. 
%\commentt{Citations here.}
%This is not feasible here.
%The extent of the array is significantly larger in $x$ than in $y$ (see Fig.~\ref{fig:modelsystem}) and the wave function of one cavity cannot consistently be distinguished from the other.
%Thus the approach using the Husimi function is more straight-forward, as the boundaries of each cavity are clearly defined.

\begin{figure*}
\centering
\includegraphics[width=.8\textwidth]{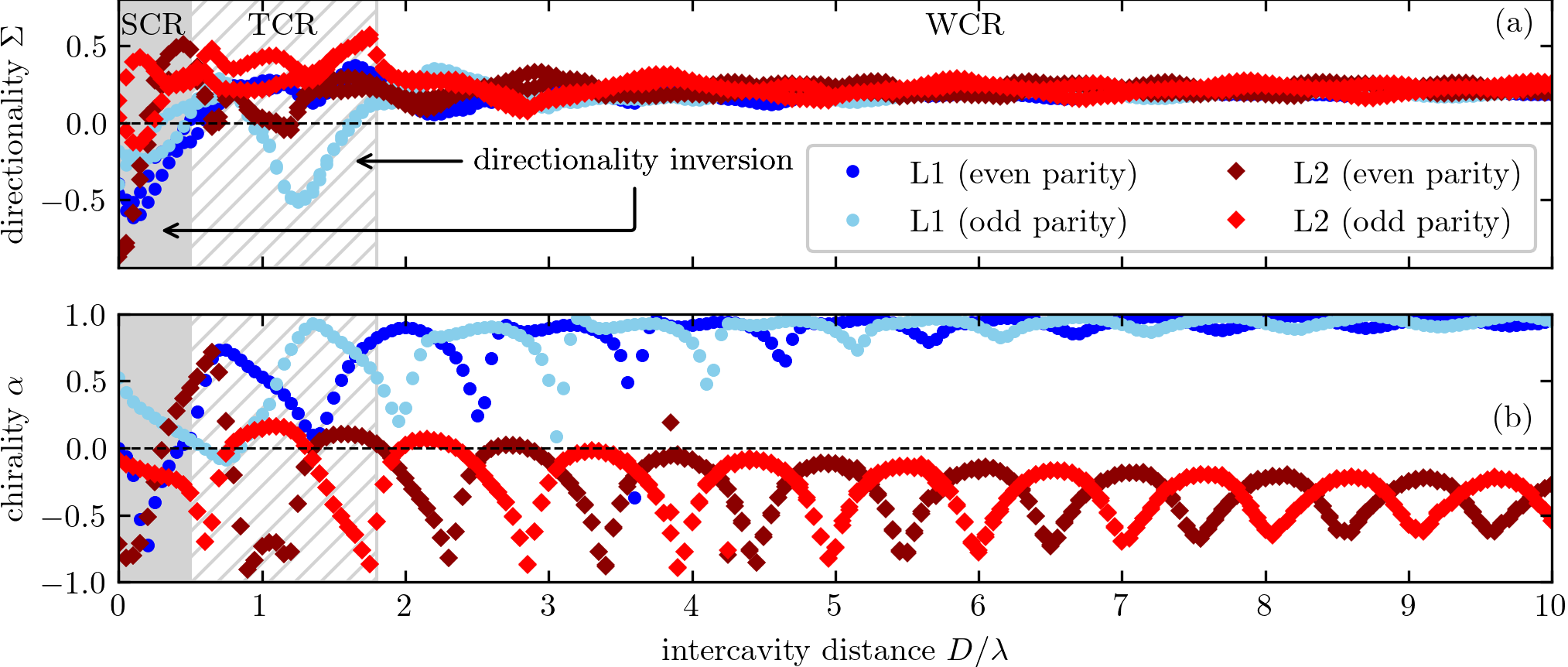}
\caption{
	\resubm{(a) Far-field directionality $\Sigma$ and (b) mode chirality $\alpha$ versus intercavity distance $D/\lambda$ for four different modes. 
		The gray (hatched, white) region indicates the strong (transient, weak) CR. While the chirality shows a strong mode dependence depending on microscopic details, the far-field emission characteristics are universal in the weak CR.
		% CHECK: and reflects the single \lima properties for intercavity distances $D/\lambda > 2$.
		% TSR: It doesn't the cavities are still too close together (see far-field figure in SM)
		For smaller distances, the transition from the oscillatory behavior to the SCR sets in, allowing for a higher $\Sigma$ or a reversal of the emission directionality (indicated by arrows) as typical non-universal effects induced by the non-Hermitian coupling at small to intermediate intercavity distances.}   
	%    %background shows the SCR, the white background the WCR (not noted). 
	%    In the whole weak CR, the chirality oscillates with decreasing amplitude as 
	%    $D/\lambda \rightarrow \infty$.  
	%    %of the modes shows an oscillating behavior in the WCR. 
	%    The two (unresolved) modes belonging to each even/odd symmetry possess
	%    %have parity, frequency, and distance have 
	%    almost identical $\alpha$. 
	%    The dependence of the (average) chirality on the resonance considered is evident.
	%    %The values for both frequencies appear to converge to a different value for %$D/\lambda \rightarrow \infty$.
	%    %The values for even and odd modes are shifted by $\Delta D/\lambda = 0.5$. 
	%    %In the SCR the oscillation ceases.
}
\label{fig:dir_and_chir}
\end{figure*}

Figure \ref{fig:dir_and_chir} shows the \resubm{far-field directionality $\Sigma$ and the} chirality $\alpha$ as a function of the scaled intercavity distance $D/\lambda$ for four different modes, namely the symmetric and antisymmetric %, SL1- and SL2-based 
coupled-cavity modes CL1 and CL2.
It reveals a large variability of the chirality that strongly depends on the underlying single cavity mode (setting a "base value"). 
%\resubm{We consider that the larger  variability of the far-field emission of the \lima cavity with $n=2.5$ when compared with the often considered $n>3$ being one reason for this, after translation in mode-dependent coupling properties (cf.~Fig.~4 of SM) 
\comments{I don't understand what you mean here}. 
\commentm{Solution: I have taken out this obviously confusing sentence (not in in this version any more!).}
\resubm{Furthermore, the chirality}
%and (ii) 
noticeably oscillates with intercavity distance $D/\lambda$, very similar to what we observed in Fig.~\ref{fig:array_eigenvalues} above. 
%RESUBM TAKEN OUT 
%Notice that the oscillations in the chirality are related to the oscillations in the eigenvalues in Fig.~\ref{fig:array_eigenvalues}.
However, the extrema do not exactly match, indicating a more intricate coupling mechanism that we  address below. %in the next section.

\resubm{The relation between chirality and far-field properties is important for applications and shown in Fig.~\ref{fig:dir_and_chir}(a). We recover a universal far-field directionality in the weak CR. Whereas the small distances required to reach the strong CR might be hard to realize in experiments, the transient CR provides a window where the far-field varies strongly and is mode dependent with $D/\lambda$, allowing the desired properties to be tailored. In particular, we observe a directionality reversal as reported in Ref.~\cite{coupled_PRR2018} albeit for more than three coupled \lima cavities in the earlier work.} 
%Taking into account the non-isotropic, directional emission characteristics of the \lima cavity, it is evident that the chirality has a strong influence on the far-field emission properties. Such an effect, including emission reversal, was previously observed in asymmetric cavities \cite{sarma_ge_wiersig} and linear cavity arrays \cite{coupled_PRR2018}.

%\section{Effective non-Hermitian Hamiltonian}
%\label{sec_anal}

\begin{figure} \label{eq:eigenvects_cmb}
\centering
\includegraphics{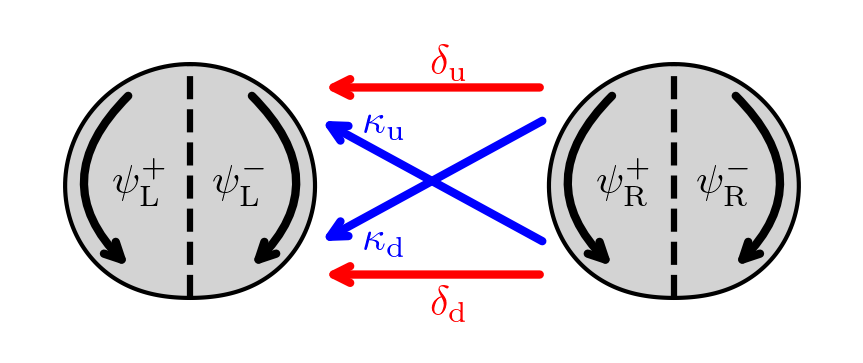}
\caption{
	Traveling-wave basis (TWB) and the most important coupling channels. 
	%as in eq.~\ref{eq:superposition_twb}. 
	%Each cavity contributes two modes $\psi^\pm$ to the system, which couple in an angular momentum changing (blue $\xcoupl$, see eq.~\ref{eq:coupl_x}) or preserving (red $\ocoupl$, see eq.~\ref{eq:coupl_o}) manner. 
	%Eigenvalue shift and internal back-scattering are not shown (see eqs.~\ref{eq:coupl_shift} and \ref{eq:coupl_backsc}).
}
\label{fig:travelling_wave_basis}
\end{figure}

\sectiontitle{Effective non-Hermitian Hamiltonian}
We will now demonstrate that our findings, namely the occurrence of eigenmodes in symmetric/antisymmetric pairs with an intrinsic, mode-dependent chirality, %that sensitively depends on the specific mode properties, 
can be understood %explained 
analytically %in a semi-quantitative manner 
using a $4\times4$ Hamiltonian. % operator. 
%\subsection{Travelling-wave basis}
The effective Hamiltonian of a coupled microcavity system can be written as
\begin{equation}
H = H_0 + H_1 \quad ,
\end{equation}
where the $4\times4$ matrix $H_0$ describes the uncoupled system (two single cavities, L(eft) and R(ight), with CCW (+) and CW (-) traveling waves) with basis $\psi^\pm_\mathrm{L/R}$ (see Fig.~\ref{fig:travelling_wave_basis}).
The intercavitiy coupling is described by $H_1$.
We use the traveling-wave basis (TWB) and decompose a wave $\Psi$ as
%. A vector $\Psi$ in the TWB can be written as
%
\begin{equation} \label{eq:superposition_twb}
\Psi^{\{\mathrm{TWB}\}} = a_1 \psi^+_\mathrm{L} + a_2 \psi^-_\mathrm{L} + a_3 \psi^+_\mathrm{R} + a_4 \psi^-_\mathrm{R} \quad .
\end{equation}
%
%where $\psi^{+/Cw}\mathrm{L/R}$ are the counter-clockwise/clockwise (CCW/CW) running WGMs in the left and right cavity, respectively. 
%Thus $H$ is a $4 \times 4$ matrix, as each cavity contributes two modes to the system.
Neglecting CCW-CW scattering, as is justified from our numerical wave simulations, we can express the uncoupled system as $H_0 = \mathrm{diag}(E_0, E_0, E_0, E_0)$ in the TWB, and set $E_0$ to zero in the following without loss of generality. 
%In the TWB the uncoupled system is described by
%
%\begin{equation}
%    H_0 = \begin{pmatrix}
%        E_0 & S & 0 & 0 \\
%        S & E_0 & 0 & 0 \\
%        0 & 0 & E_0 & S \\
%        0 & 0 & S & E_0
%    \end{pmatrix} \quad,
%\end{equation}
%
%where a single cavity is represented by a $2 \times 2$ block on the main diagonal with mode eigenenergy $E_0$.
%$S$ represents the backscattering between CW and CCW traveling waves %due to the cavity deformation. %\commentm{What about the other way round, it is actually not clear that this yields the same S!!! }
%Wave simulations justify the approximation $S \approx 0$.
%due to the \lima only deviating slightly from the disc shape. % (see Fig.~\ref{fig:modelsystem}).
%NOTE MH: THis slight deviation was enough to induce heavy chiralities! 

The coupling matrix $H_1$ has several contributions, including an overall shift of the resonance energies and coupling-induced scattering between CCW and CW components. Most importantly, the \lima symmetry implies that coupling in the upper ``u'' part differs from the one in the lower or down ``d'' part, cf.~Fig.~\ref{fig:travelling_wave_basis} where the most important coupling mechanisms are indicated.
Denoting the intercavity coupling between the same sense of rotation in the left and right cavity, $\psi_{\mathrm{L/R}}^+ \rightarrow \psi_{\mathrm{R/L}}^+$ or $\psi_{\mathrm{L/R}}^- \rightarrow \psi_{\mathrm{R/L}}^-$ by $\delta_{u}$ and $\delta_{d}$; and the coupling between opposite senses of rotation, $\psi_{\mathrm{L/R}}^+ \rightarrow \psi_{\mathrm{R/L}}^-$ or $\psi_{\mathrm{L/R}}^- \rightarrow \psi_{\mathrm{R/L}}^+$ by $\kappa_{u}$ and $\kappa_{d}$, the coupling matrix $H_1$ can, in the weak CR at larger $D/\lambda$, approximately be written as 
\begin{equation} \label{eq:coupl_x}
H_1 = H = \begin{pmatrix}
	0 & 0 & \ocouplu & \xcouplu \\
	0 & 0 & \xcoupld & \ocoupld \\
	\ocoupld & \xcoupld & 0 & 0 \\
	\xcouplu & \ocouplu & 0 & 0 \\
\end{pmatrix} \quad .
\end{equation}

\sectiontitle{Composite wave basis.}
Instead of solving directly for the eigenvalues and eigenmodes of $H$, a transformation to the composite wave basis (CWB) simplifies things considerably. We introduce the CWB basis states as 
%General solutions of $4 \times 4$ matrices give very complex results, which are very hard to intuit.
%To arrive at more straight-forward solutions the composite-mode basis (CMB) with
%
%\begin{equation}
%    \Psi^{\{CMB\}} = b_0 \psi^+_\mathrm{s} + b_1 \psi^-_\mathrm{s} + b_2 \psi^+_\mathrm{a} + b_3 \psi^-_\mathrm{a}
%\end{equation}
%
%is introduced. The basis vectors
%
\begin{subequations}
\label{eq:basisvects_cmb}
\begin{align}
\label{eq:basisvects_cmb_symm}
\psi^\pm_\mathrm{s} &= (\psi^\pm_\mathrm{L} + \psi^\mp_\mathrm{R}) / \sqrt{2} \quad \text{and} \\
\label{eq:basisvects_cmb_antisymm}
\psi^\pm_\mathrm{a} &= (\psi^\pm_\mathrm{L} - \psi^\mp_\mathrm{R}) / \sqrt{2} \quad ,
\end{align}
\end{subequations}
i.e. as symmetric and antisymmetric "figure-of-eight"-modes that capture, or anticipate, the coupled system's properties right away. 
%with normalization factor $1/\sqrt{2}$ are symmetric/antisymmetric superpositions of TWB eigenvectors (denoted by the subscribts s/a).
%They reflect the mirror symmetry of the system.
The superscripts $\pm$ denote the angular momentum in the left cavity.

Note that the CWB is equivalent to the standing-wave basis (SWB), as a vector in SWB can always be transformed to TWB using the transformation matrix $M$
\begin{equation}
M = \frac{1}{\sqrt{2}} \begin{pmatrix}
1 & 0 & 1 & 0 \\
0 & 1 & 0 & 1 \\
0 & 1 & 0 & -1 \\
1 & 0 & -1 & 0
\end{pmatrix}
\end{equation}
with $\Psi^{\{\mathrm{CWB}\}} = M \Psi^{\{\mathrm{TWB}\}}$ and $M^{-1} = M^\mathrm{T}$.
An expression of $H_1$ in CWB is obtained by the similarity transformation
\begin{equation} 
\nonumber
\label{eq:hamiltonian_cmb}
H_1^{\{CWB\}} = M H_1^{\{TWB\}} M^\mathrm{T} = 
\begin{pmatrix}
%   A + B & 0 \\
%    0 & A - B
B & 0 \\
0 &  - B
\end{pmatrix} \:; \quad 
B = \begin{pmatrix}
\xcouplu & \ocouplu \\
\ocoupld & \xcoupld
\end{pmatrix} \: .
\end{equation}
%
%with
%
%\begin{equation}
%%    A = \begin{pmatrix}
%%        \evshiftu & \backscu \\
%%        \backscd & \evshiftd
%%    \end{pmatrix} 
%%    \quad \mathrm{and} \quad 
%    B = \begin{pmatrix}
%        \xcouplu & \ocouplu \\
%        \ocoupld & \xcoupld
%    \end{pmatrix} \quad .
%\end{equation}
%
%and
%
%\begin{equation}
%    B = \begin{pmatrix}
%        \xcouplu & \ocouplu \\
%        \ocoupld & \xcoupld
%    \end{pmatrix} \quad .
%\end{equation}
%
Remarkably, this reduces the original $4 \times 4$ Hamiltonian $H$ 
%in eq.~\ref{eq:hamiltonian_twb} 
to two uncoupled $2\times2$ Hamiltonians $B$ with opposite signs in the weak CR.
%Further terms arise in the SCR and will be discussed elsewhere. 
%in eq.~\ref{eq:hamiltonian_cmb}. All perturbation of the optical potential of one cavity is given by $A$, while $B$ describes scattering processes of the CW/CCW modes between the cavities.

%In the WCR the perturbation of the optical potential of a cavity is %negligible. It follows that $A \approx 0$. Due to the two parities %being decoupled, we focus on the even modes which are described by $B$. The eigensystem of $B$ gives the eigenvalues
%
It can be readily solved and yields the eigenvalues
\begin{equation} \label{eq:eigvals_cmb}
E_{1,2} = (\xcouplu + \xcoupld) / 2 \pm \sqrt{(\xcouplu - \xcoupld)^2 / 4 + \ocouplu \ocoupld}
\end{equation}
and eigenvectors
%
%\begin{equation} \label{eq:eigvects_cmb}
%    \Psi_{1,2} = \begin{pmatrix}
%        (\xcouplu - \xcoupld)/2 \pm \sqrt{ (\xcouplu-\xcoupld)^2/4 + \ocouplu \ocoupld } \\
%        \ocoupld
%    \end{pmatrix} \quad .
%\end{equation}
%
\begin{equation} \label{eq:eigvects_cmb}
\Psi_{1,2} = \begin{pmatrix}
C_{1,2} \\
\ocoupld
\end{pmatrix} \quad ,
\end{equation}
with 
%\commentt{$\pm \rightarrow \mp$ for $C_{1,2}$ comes from analytical solution, I noticed this just now. This has no impact on our results}
%
%\begin{align}
\begin{equation}
C_{1,2} = (\xcouplu - \xcoupld)/2 \mp \sqrt{ (\xcouplu-\xcoupld)^2/4 + \ocouplu \ocoupld } \quad,
\end{equation}
%    C_{1,2} &= (\xcouplu - \xcoupld)/2 \pm \sqrt{ (\xcouplu-\xcoupld)^2/4 + \ocouplu \ocoupld } &\quad \text{and} \\
%   D       &= \ocoupld &\quad .
%\end{align}
%\resubm{Note that there are several ways in which the eigenvectors can be expressed. This expression assumes 
\resubm{where we have, without loss of generality, assumed $\ocoupld \neq 0$. If this condition is not fulfilled, "up" and "down" (that were chosen arbitrarily) need to be switched, thus %the values 
$\xcoupld \leftrightarrow \xcouplu$ and $\ocoupld \leftrightarrow \ocouplu$.}
%need to be switched. This is reflecting the fact that "up" and "down" are chosen arbitrarily.} 
This explains the principal findings of Fig.~\ref{fig:array_eigenvalues}, namely that the eigenvalues of the coupled system occur in pairs (1,2) with opposite signs, i.e., opposite deviations from the single cavity value.
As seen in Fig.~\ref{fig:array_eigenvalues}, the values $E_{1,2}$ nearly coincide, implying $\xcouplu \approx \xcoupld$.

%\subsection{Exceptional points and limiting cases}
\sectiontitle{Limiting cases and exceptional points.}
% \commentt{Here I think the Husimi function measures the SQUARE of the modulus, but the derivation works the same. Please check!}
% \resubm{CHECK CONDITIONS FOR EP REFEREEE 2.}
The upper and lower components of the eigenvector in Eq.~(\ref{eq:eigvects_cmb}) are a measure for the CW/CCW weight, respectively.
This yields the approximate  chirality in the CWB, cf.~also Ref.~\onlinecite{nonorth_spiral},
%with eq.~\ref{eq:chirality}, we arrive at
%As Eq.~(\ref{eq:eigvects_cmb}) contains all information about CCW and CW components, it allows us to rewrite the chirality $\alpha$ as
%allows as
%Emerging mode chirality is a key indicator of the system being close to an exceptional point. 
%How chirality from eq.~\ref{eq:chirality} relates to the CWB solutions in eqs.~\ref{eq:eigvals_cmb} and \ref{eq:eigvects_cmb} will be determined in the following. 
%The chirality can also be written as \commentm{WHY}
%
\begin{equation}
\tilde{\alpha} =
%    S = \frac{\Psi_1^* \Psi_2}{|\Psi_1| %\times 
%    |\Psi_2|} = 
\frac{1 - | \ocouplu / \ocoupld |}{1 + | \ocouplu / \ocoupld |} \quad ,
\label{eq:chirality_cmb}
\end{equation}

%\begin{equation} \label{eq:chirality_cmb}
%    \tilde{\alpha}_{1,2} = \frac{|C_{1,2}| - |\ocoupld |}{|C_{1,2}| + |\ocoupld |} \quad .
%\end{equation}
%
%because the Husimi function measures the absolute values of the contributions. 
%\commentm{verstehe folgenden Satz nicht!}
%\commentt{Now better?}
%Making use of $\xcouplu \approx \xcoupld$, we find
%Considering the nearly identical values within even/odd modes in Fig.~\ref{fig:dir_and_chir}, $\alpha_1 \approx \alpha_2$ is evident.
%This implies $\xcouplu \approx \xcoupld$, which allows one to
%Thus the chirality can be estimated to be
%
%
%Making use of $\xcouplu \approx \xcoupld$, we find
%\begin{equation} \label{eq:chirality_ocouplratio}
%    \tilde{\alpha} = \tilde{\alpha}_1 = \tilde{\alpha}_2 = \frac{|\sqrt{\ocouplu / \ocoupld}| - 1}{|\sqrt{\ocouplu / \ocoupld}| + 1} \quad ,
%\end{equation}
%
implying that the chirality only depends on the ratio of the two angular-momentum-changing coupling parameters $\delta_{u,d}$. 
This also explains the pronounced chirality found even for large distances in the weak CR where the energy-level splitting is very small.
%(compare WCR in figs.~\ref{fig:array_eigenvalues} and \ref{fig:dir_and_chir}). 
The ratio of $\ocouplu$ and $\ocoupld$ can take on large values, even if both of them are small. 
%It is evident that $|\ocouplu| > |\ocoupld|$ for positive $\alpha$ and the contrary for negative $\alpha$ are prevalent.
At the same time, vanishing chirality for the two coupled disks is evident as $\xcouplu = \xcoupld$ and $\ocouplu = \ocoupld$, in agreement with our previous findings. 
%5For the disc array the chirality is $\alpha = 0$ for all $D/\lambda$. 
%This can be seen in the symmetry of the Husimi function in Fig.~\ref{fig:husimi_disc}. 
%Due to the symmetry along the $x$-axis $\xcouplu = \xcoupld$ and $\ocouplu = \ocoupld$ are enforced trivially.

Emerging mode chirality is a key indicator for a non-Hermitian system being close to an exceptional point where degenerate eigenvalues and -vectors 
%At an EP the eigenvalues and eigenvectors of two modes 
%are degenerate, 
result in $|\tilde{\alpha}| = 1$ \cite{Bender_2007, nonorth_spiral,shim_kim_expoint2009}.
This %can be 
is achieved  setting %one of 
$\ocoupld$ or $\ocouplu$ to zero \resubm{while $\xcoupld = \xcouplu$}. 
To what extent this condition might be fulfilled by some specific resonances, or can be enforced in experiments \resubm{or realized in other structures\cite{taiji-PhysRevA.104.043501}} will be the subject of further studies.  
%Experimentally, absorbing or reflecting material could be introduced into one of the two coupling regions, so that only one of the two mechanisms remains.
%Note that the distinction of ``up'' and ``down'' used here is arbitrary.
%This means that $\alpha \rightarrow \infty$ for $\ocoupld = 0$ in eq.~\ref{eq:chirality_ocouplratio} can be lifted by a rotation of the system, which results in a switch of the definitions for $\ocouplu$ and $\ocoupld$.

%\section{Conclusion}

\sectiontitle{Conclusion.}
\resubm{To summarize,}
we have investigated a system of two coupled non-circular microdisks of \lima shape. 
We have illustrated that the reduced symmetry of \resubm{(any)} deformed microdisks introduces non-Hermitian physics into the coupled system, immediately visible via the chirality of the modes of the coupled system that does not, in contrast, occur for coupled ideal disks. 
We are able to capture the essential features in an intuitive analytic model. 
%RESUBM TAKEN OUT 
%that reveals the underlying non-Hermitian physics properties, i.e, the fact that the sign and extent of the chirality are both mode and intercavity distance dependent.
%\comments{I rewrote this sentence as it wasn't clear before - please check it}
%The identification of exceptional points that ultimately rule the non-Hermitian Hamiltonian, and the relation between chirality and the far-field emission properties of coupled systems, will be the subject of further work.
There is an intimate relation between chirality and emission properties that we illustrated in Fig.~\ref{fig:array_modes} \resubm{and that opens the possibility to fine-tune the near and midfield emission properties via the intercavity distance $D/\lambda$. The far-field emission depends sensitively on $D/\lambda$ especially in the transient CR for $D$ of the order of $\lambda$ and suggests the possibility of versatile far-field control, including an emission reversal,} by just a slight change of the coupling interaction via the intercavity distance.
%, as well as the study of more than two coupled microdisks.

%\section{Acknowledgments}

\sectiontitle{Acknowledgements.}
S. Nic Chormaic acknowledges support from the Technische Universität Chemnitz Visiting Scholar Program and from the Okinawa Institute of Science and Technology Graduate University.
The authors thank R. Madugani, S. Li, J. Kullig, and J. Wiersig for fruitful discussions.

%\begin{itemize}
%    \item summary
%    \item notice that the results in ref. [Kreismann] were obtained %in meep, where more than two cavities were excited
%    \item outlook
%    \begin{itemize}
%        \item more than two cavities
%        \item exceptional points
%        \item deeper far-field analysis with antenna theory (cut here %completely due to page number constraints)
%        \item relationship to Kreismann?
%    \end{itemize}
%    
%\end{itemize}

%\section*{Data availability and Author declarations}

\sectiontitle{Data availability and Author declarations.}
The data that support the findings of this study are available from the corresponding author upon reasonable request.
The authors have no conflict of interest to disclose.

\bibliography{literatur}

\end{document}

\maketitle

\section{Eigenvalue behavior}

\begin{figure}[!b]
	\centering
	\includegraphics[width=0.45\textwidth]{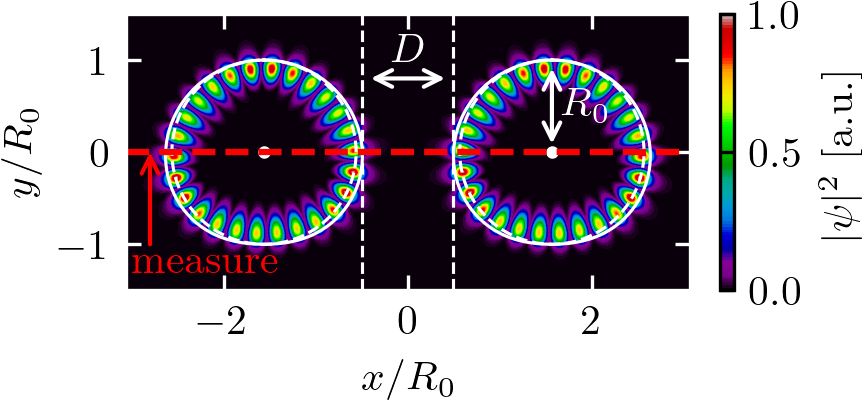}
	\caption{
		Model system consisting of two \lima microcavities. The even (with respect to the intercavity axis) CL1 mode intensity distribution is shown at an intercavity distance $D/\lambda=1$.
		The red, dashed line marks the path along which the field intensities are measured for Fig.~\ref{supp:fig:slice_intensities}.
	}
	\label{supp:fig:slice_line}
\end{figure}

\begin{figure*}
	\centering
	\includegraphics[width=0.95\textwidth]{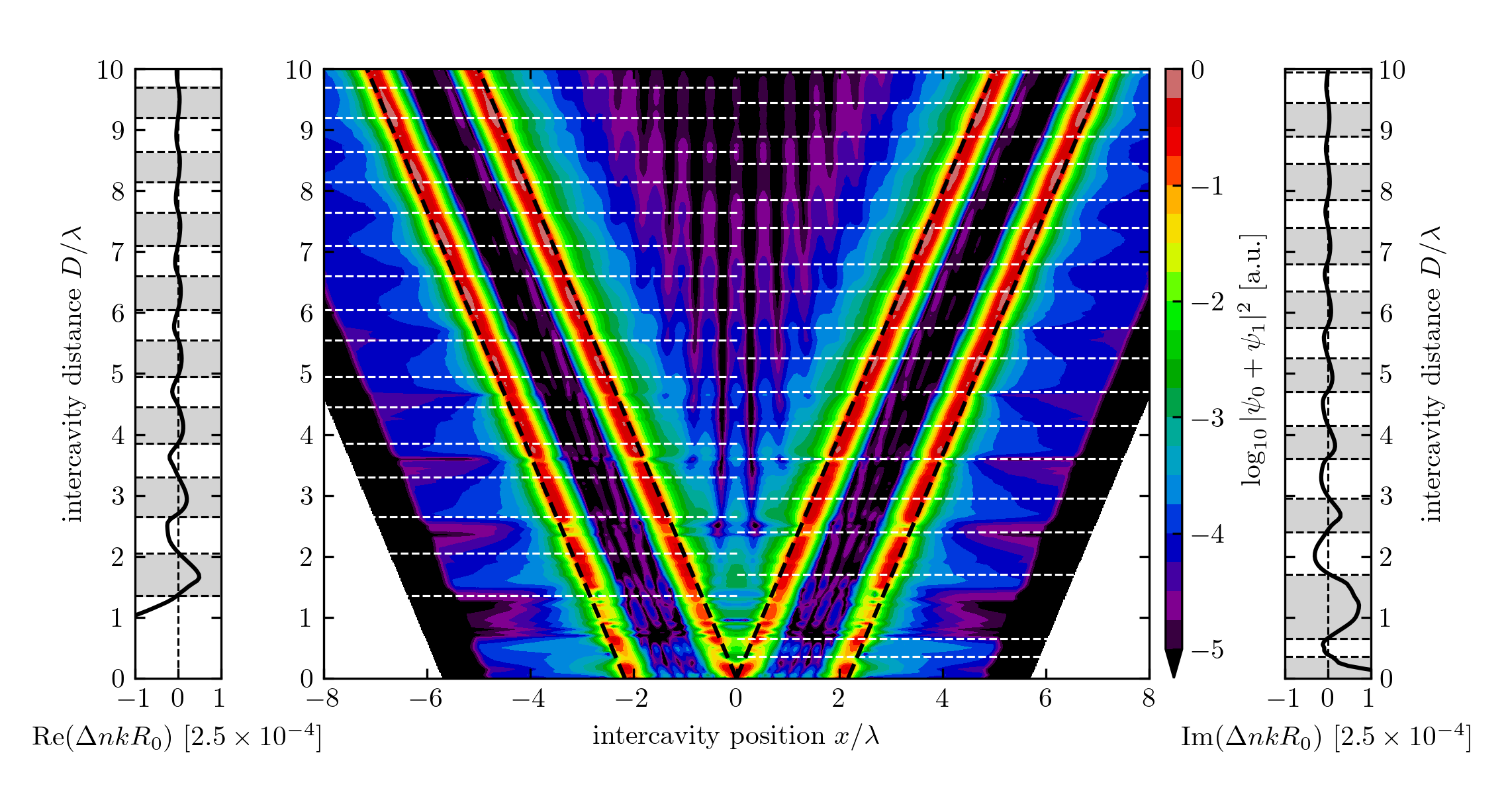}
	\vspace{-2em}
	\caption{
		Central plot: Logarithmic electromagnetic field intensity of the even CL1 mode (Fig.~\ref{supp:fig:slice_line},  spatial position $x/\lambda$ taken along the red line) as a function of the intercavity distance $D/\lambda$, averaged over the two possible intracavity mode symmetries. 
		The black dashed lines mark the boundaries of the two cavities.
		Left (right) plot: real (imaginary) part of deviation from the SL1 eigenfrequency $nkR_0$.
		%The intercavity position $x/\lambda$ corresponds to the location in the system (see red line in Fig.~\ref{supp:fig:slice_line}). 
		The horizontal dashed lines in the plots for $\Delta n k R_0$ correspond to the zeros. Where the deviation is positive (negative), the background is gray (white). The white lines in the center indicate distances $D/\lambda$ where $\mathrm{Re}(\Delta nkR_0)=0$ and $\mathrm{Im}(\Delta nkR_0)=0$, respectively, and guide the eye through the change in the mode pattern.
		%plot are extensions of the zero lines in the $nkR_0$ plots. They demonstrate how changes in $\Delta nkR_0$ correspond to a shift in mode pattern. 
		The white areas near the lower corners %are coordinates without data, as they 
		lie outside the computational cell. The adjacent thick, black areas %adjacent to the coordinates without data 
		correspond to the perfectly matched layers (PMLs) surrounding the system. %cell. 
		% The field is very low there, as the absorbing properties of the PMLs model the infinite space in which the system is placed.
	}
	\label{supp:fig:slice_intensities}
\end{figure*}

We first illustrate the evolution of the eigenvalues of the coupled system as the interacvity distance $D$ is varied. Our focus is to show how the quantization condition along the intercavity symmetry axis (along $x=0$) is realized by (slightly) adjusting the real and imaginary part of the eigenvalue to a given $D$. To this end, 
%In order to gain a fundamental understanding of the influence of the coupling we analyze the field intensities between the two cavities. For this, 
we measure the electric field intensity along a line connecting the (effective) centers of our \lima cavities,
%of the equivalent discs of each cavity (
indicated by the red line in Fig.~\ref{supp:fig:slice_line}). 

The behavior of the eigenvalues is shown in Fig.~\ref{supp:fig:slice_intensities}. As the field intensities can be very low in between the cavities, a logarithmic scale is used.
Generally the intensity is highest inside the cavities near the boundary, as is characteristic for whispering gallery (WG) type modes. Outside the cavity, the evanescent field decays exponentially for about half a wavelength, whereas the decay becomes much weaker for larger distances. This slower decay is the origin of radiation losses, and provides the basis for coupling even at distances of several wavelengths.

Let us first consider the deviation of the real part (eigenfrequency) of the eigenvalue from the single cavity resonance, $\mathrm{Re}(\Delta nkR_0)$ in Fig.~\ref{supp:fig:slice_intensities}. 
%It corresponds to the energy level of the mode. 
A higher energy results in a shorter wavelength and effectively, a stronger confinement within the cavity. 
For example, for the SCR and TCR up to $D/\lambda < 1.5$ the frequency is lowered considerably and the field has a higher intensity in between the cavities. Thus, the mode exhibits less confinement. For larger separation distances, oscillatory behavior occurs. The even symmetry of the system demands an intensity maximum at $x/\lambda = 0$, so the resonant 
\comments{resonant? }
\commentm{Yes!}
wavelength adjusts to accommodate this condition 
\comments{this may lead to a query... how would an experiment permit this if things are done at fixed wavelength?}. 
\commentm{never thought about this, but is is tiny changes, within the experimental line width anyway --- I thought at least.}\comments{yes it should be, but best to be ready for possible comment back}
This results in alternating shifts between positive and negative deviations 
%\comments{deviatio????} --> fixed right away, 
as more and more intensity maxima fit in between the cavities upon increasing $D$.

The complex component $\mathrm{Im}(\Delta nkR_0)$ corresponds to the width of the frequency spectrum of the eigenmode and is related, via the $Q$-factor, to the mode confinement as well.
%(e.g. when the eigenfrequency is measured by coupling to a waveguide).
The emergence of new intensity maxima along the intercavity symmetry axis is enabled by (slightly) decreasing the wavelength, thereby increasing the mode confinement ($\mathrm{Im}(\Delta nkR_0) <0$), such that a new maximum can be formed around $x=0$. As the distance $D$ is (further) increased, the wavelength relaxes to its single cavity (base) value and increases beyond this, implying a decrease in the mode confinement ($\mathrm{Im}(\Delta nkR_0) > 0$). The precise, interwoven sequence is shown in Fig.\ref{supp:fig:slice_intensities}.

%\commentm{COULD NOT UNDERSTAND AND TOOK OUT:}
%When the intercavity distance is increased and new intensity maxima between the cavities emerge, at a %specific distance the maxima are their broadest. 
%%See, for example, the maximum of $\mathrm{Im}(\Delta nkR_0)$ at $D/\lambda \approx 2.75$, where the %intensity maxima are broader than for adjacent \comments{slight different rather than adjacent?} distances.
%At such intercavity distances, $nkR_0$ adjacent to the main frequency "fit" better into the system, leading to a broader frequency distribution and thus a higher $\mathrm{Im}(\Delta nkR_0)$.

\section{Influence of chirality on the far-field spectrum}

\begin{figure}
	\centering
	\includegraphics[width=.48\textwidth]{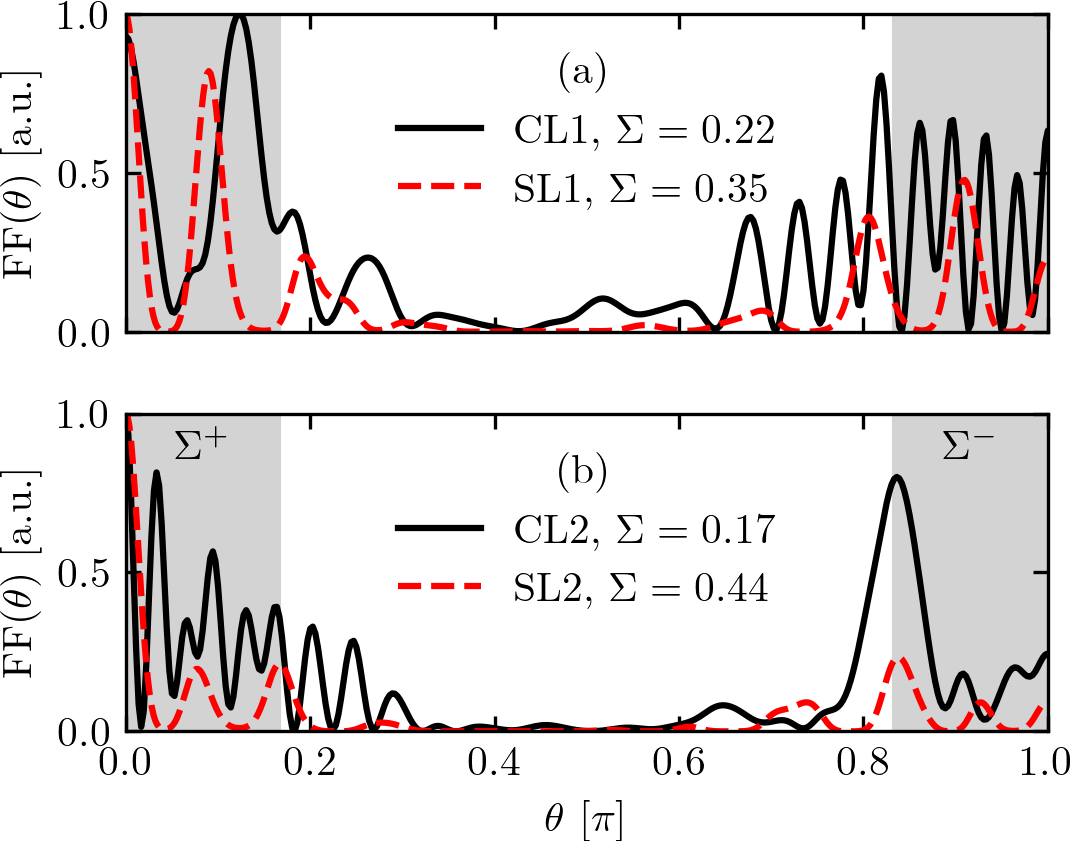}
	\caption{Far-field spectra $\mathrm{FF}(\theta)$ and directionalities $\Sigma$ of two symmetric modes at $D/\lambda = 3.4$. Only the spectrum for $\theta \in \left[0,\pi\right]$ is shown, the rest is given by symmetry. SL1/CL1 is depicted in the upper, SL2/CL2 in the lower part of the figure. The black, solid lines corresponds to the far-field emission of the array. The red, dashed lines show the far-field of the single cavity. The array modes shown here are the same as in Fig.~4. The sections with gray backgrounds mark the contributions to the far-field directionality $\Sigma$, with $\mathrm{FF}(\theta \approx 0)$ (and $\mathrm{FF}(\theta \approx \pi)$) contributing to $\Sigma^+$ (and $\Sigma^-$).}
	\label{supp:fig:farfield}
\end{figure}

A useful quantity when discussing microcavities is their far-field directionality. It is defined as
\begin{equation}
	\Sigma = \frac{\Sigma^+ - \Sigma^-}{\Sigma^+ + \Sigma^-}
\end{equation}
with contributions
\begin{subequations}
	\begin{eqnarray}
		\Sigma^+ &= \int_{-\pi /6}^{+\pi/6} \mathrm{FF}(\theta) \, \mathrm{d}\theta &\quad \text{and} \\
		\Sigma^- &= \int_{\pi 5/6}^{\pi 7/6} \mathrm{FF}(\theta) \, \mathrm{d}\theta &\quad \text{,}
	\end{eqnarray}
\end{subequations}
where $\mathrm{FF}(\theta)$ is the far-field spectrum of any given mode.\cite{coupled_PRR2018} The angles contributing to $\Sigma^\pm$ are shown in Fig.~\ref{supp:fig:farfield}. The high directionality of the \lima (see red dashed lines) is one of its key attributes when considering applications.

The chirality that emerges by coupling the microcavities can be deduced from their far-field spectra $\mathrm{FF}(\theta)$. An uncoupled \lima microcavity has two main emission centers, located at $s_1 \approx 0.3$ and $s_2 \approx 0.7 $. These can be seen as maxima in the Husimi function in Fig.~1(b). In the uncoupled case, both centers radiate in the forward ($+$, for $\theta \in \left[-\pi/2, \pi/2\right]$) and the backward ($-$, for $\theta \in \left[\pi/2, 3\pi/2\right]$) direction. The radiation of both centers toward far-field angles $\theta=0$ and $\theta = \pi$ is equal due to the balance of CW and CCW modes (see Fig.~1(b).

When chirality is introduced, CW and CCW modes are no longer balanced. Consider the left cavity in Fig.~4(a). Due to the dominance of CCW components, the emission center at $s_1$ ($s_2$) now radiates predominantly in the $-$ ($+$) direction. Due to the symmetry of the system, the two "antennae" emitting in the $+$ direction are much closer together than their counterparts emitting in the $-$ direction. The interference of these antennae can be written as a superposition \cite{balanis2016antenna} 
%\comments{check reference } --> done and taken out.
of the individual far-fields of the emission centers $\mathrm{FF}_j(\theta)$ with
\begin{equation}\label{supp:eq:antenna}
	\mathrm{FF}(\theta) = \sum_j \mathrm{FF}_j(\theta) \, \mathrm{e}^{- \imag (j-1) k D \sin \theta} \quad .
\end{equation}
The exponential function represents the phase shift emerging between the emission centers due to different path lengths $(j-1) D \sin \theta$ occurring at each far-field angle $\theta$. $k$ is the vacuum wave number. $kD$ can be understood as an oscillation frequency in $\sin \theta$ which increases with the distance between the emission centers $D$.

The far-field spectra corresponding to the two systems in Fig.~4 are shown in Fig.~\ref{supp:fig:farfield}. Let us first examine CL1. Figure 4(a) shows that the distance between the two emission centers in the forward direction is significantly smaller than the distance of the backward-facing centers. This would suggest a lower (higher) modulation-frequency for $\Sigma^+$ ($\Sigma^-$) which is clearly evident in Fig.~\ref{supp:fig:farfield}(a). Also note that the far-field spectra (and also the directionality) of SL1 and CL1 are very different, even though the cavities are in the weak-coupling regime.

Due to the inverse chirality of the CL2 mode, we expect the opposite to be the case. The emission center distances in Fig.~4(b) are emitting in opposite directions in comparison to CL1. This is shown in Fig. \ref{supp:fig:farfield}(b), where a faster (slower) oscillation in $\theta$ for $\Sigma^+$ ($\Sigma^-$) occurs.

\section{Chirality of two additional modes and numerical stability}

\begin{figure*}
	\centering
	\includegraphics{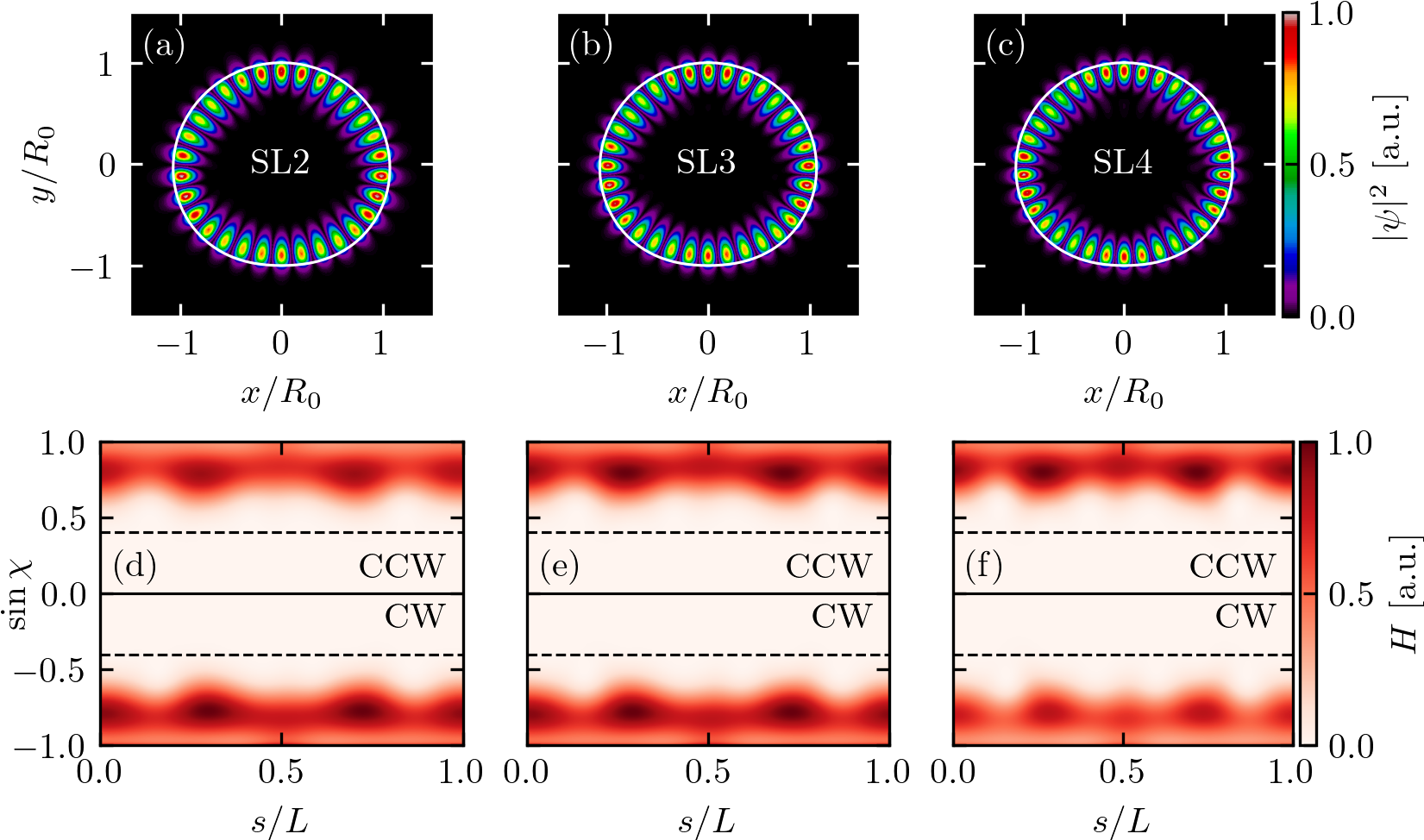}
	\caption{
		(a-c) Mode pattern and (d-f) inner, incident Husimi functions of the (a,d) SL2, (b,e) SL3 and (c,f) SL4 modes.
		The white line represents the cavity boundary.
		The dashed line in (b,d,f) indicates the sine of the critical angle $|\sin \chi_\mathrm{crit}| = 1 / n$ with positive (negative) $\sin \chi$ indicating counterclockwise, CCW (clockwise, CW) motion.
		The Husimi function shows the WG mode characteristics as it is confined to the region with total internal reflection outside the critical lines.
	}
	\label{supp:fig:new_modes}
\end{figure*}

\begin{figure*}
	\centering
	\includegraphics[width=.95\textwidth]{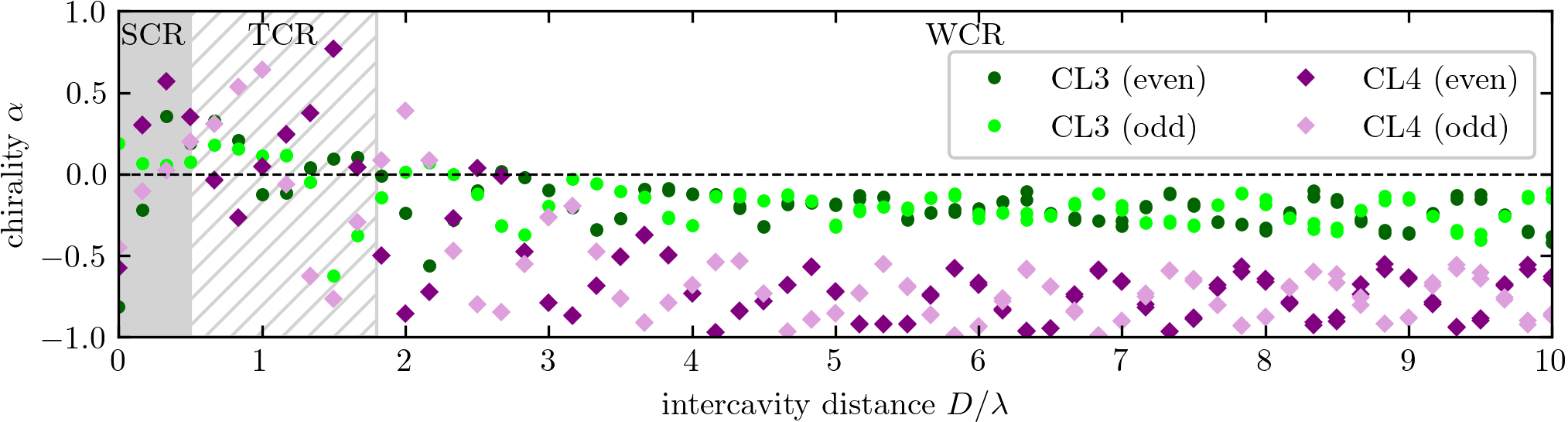}
	\caption{Chiralities of CL3 and CL4 modes as a function of intercavity distance $D/\lambda$. The gray (hatched, white) region corresponds to the SCR (TCR,WCR). We confirm the realization of a wide range of chirality values that oscillate as afunction of diatnce $D/\lambda$ about a mode-dependent value. %chiralities oscillating , the gray shaded r \comments{seems incommplete}--> fixed.
	}
	\label{supp:fig:chiralities}
\end{figure*}

In order to demonstrate the nature of the chirality in the weak coupling regime, the results for two additional modes are shown in this section. The additional modes are SL3 with $nkR = 19.09$ ($Q = 5.05 \times 10^4$) and SL4 with $nkR = 20.15$ ($Q = 3.35 \times 10^4$). Field and Husimi functions of the unperturbed modes are shown in Fig.~\ref{supp:fig:new_modes}.  For reason of comparison, Fig.~\ref{supp:fig:new_modes} also features SL2. 
%, which has already been discussed in the main section of the article. 
Figure~\ref{supp:fig:chiralities} shows the chiralities of the coupled modes CL3 and CL4. Our previous findings are confirmed as we find the chirality to display oscillatory behavior in the WCR and %appears
to approach a constant, eigenmode-dependent value for large $D/\lambda$.
%that is distinct for each eigenmode.

%Let us examine the numerical stability of the calculations. Also note that the chirality of CL4 is generally smaller than that in the WCR. This is in contrast to the bias to the positive sense of rotation, see Fig.~\ref{supp:fig:new_modes}(f). The imbalance of CW/CCW components in the uncoupled mode is due to the calculation grid slightly breaking the symmetry. This can lead to a visible imbalance, even when the grid has a much higher resolution than is typically required to resolve a wave. Due to the relationship in Eq.~(\ref{eq:chirality}), these small breaks of symmetry can have an impact. Nevertheless the results herein are robust and cannot be explained by random grid asymmetries, since 
%\begin{itemize} 
%	\item 
\comments{don't particularly like this listing in a  paper... better to number within the text }
\commentm{have taken out this paragraph to avoid confusion (on all levels)}
%he asymmetry introduced by the other cavity dominates the small effects of the grid (compare bias of SL4 in Fig.~\ref{supp:fig:chiralities}(f) to Fig.~\ref{supp:fig:chiralities}), and
%	\item the chirality measured in both cavities differs by exactly a factor of $-1$. If grid geometry affected the measurements in a significant way, the chirality in both cavities would be independent of each other.
%\end{itemize}

\section{Phase-space analysis}

\begin{figure*}[p]
	\centering
	\includegraphics[width=1.\textwidth]{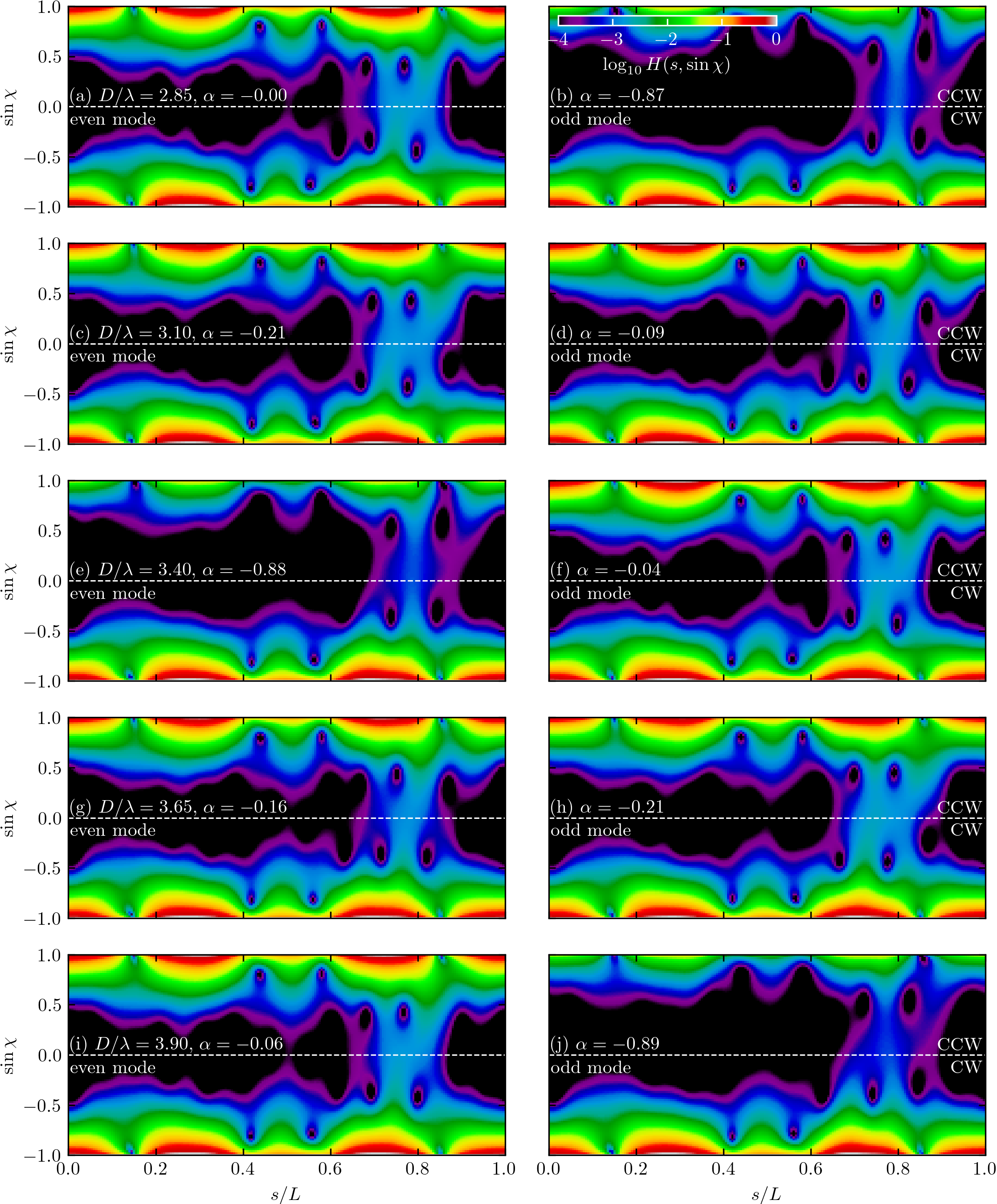}
	\caption{Outer, emerging Husimi functions (color scale) for CL2 modes with intercavity symmetry being  even (left hand side) and odd (right hand side) 
		%CL2 modes 
		for various intercavity distances  and resulting chiralities as indicated. The Husimi function is calculated for the left cavity. Note the logarithmic intensity scale, which is shown in (b). 
		%Each line depicts the Husimi function for one $D/\lambda$. 
		%The left (right) column depicts Husimi functions of even (odd) modes. 
		The distances $D/\lambda$ considerd range from one minimum chirality of the odd modes to the next. We point out the oscillation of the Husimi pattern between the even and odd mode. Note that alltogether two even modes  and two odd modes exist for each $D/\lambda$. Here, we focus on one of the two intracavity parities and report that they only slightly differ. 
		%was chosen, representative Husimi function (one intracavity parity) per parity is depicted here.
	}
	\label{supp:fig:husimi_evolution}
\end{figure*}

Another useful tool is available when we look at the the outer Husimi functions.\cite{husimiEPL} They give insight into the microscopic emission properties of any given mode. Outer Husimi functions of CL2 modes for various intercavity distances are shown in Fig.~\ref{supp:fig:husimi_evolution}. The intensity is mostly concentrated at around $\sin \chi \pm 1$, as is characteristic for WG-type modes considered here. The Husimi function increases at $s/L \approx 0.8$, which corresponds to the location of the coupling region (cavity boundary closest to the other cavity), especially for small $|\sin \chi|$.
% in the system. This behavior is  due to the coupling between the cavities.

The presence of chirality is visible in terms of different Husimi amplitudes around $\sin \chi \pm 1$. See, for example, Fig.~\ref{supp:fig:husimi_evolution}(e) where the negative sense of rotation leads to significantly higher Husimi values at $\sin \chi = -1$ in comparison to $\sin \chi = 1$. For even and odd modes of similar chirality, the Husimi functions possess very similar features. The shift between the two parities is $\lambda/2$ due to their differing symmetry conditions.

For low chiralities ($\alpha \approx 0$, see Figs.~\ref{supp:fig:husimi_evolution}(a) and (f)) a broad structure around $s/L \approx 0.8$ is present. % which roughly possesses a rotational symmetry. 
However, in the imbalanced case with large chirality (e.g. $\alpha \approx -0.9$ in Figs.~\ref{supp:fig:husimi_evolution}(b) and (e)), the size of this phase-space structure 
%in $s/L$ of the phase space structure 
shrinks. The overall weight of the Husimi function also shifts towards negative $\sin \chi$, a feature that is directly related to the preference of one sense of rotation. Deciphering the phase space structure and relating it more closely to the coupling coefficients will be the subject of future work.

\bibliography{literatur.bib}